\documentclass[sigconf, authorversion]{acmart}
\AtBeginDocument{%
  }

\copyrightyear{2026}
\acmYear{2026}
\setcopyright{cc}
\setcctype{by}
\acmConference[ICER 2026 Vol. 1]{Proceedings of the ACM Conference on International Computing Education Research Vol.1}{August 11--14, 2026}{Uppsala, Sweden}
\acmBooktitle{Proceedings of the ACM Conference on International Computing Education Research Vol.1 (ICER 2026 Vol. 1), August 11--14, 2026, Uppsala, Sweden}
\acmDOI{10.1145/3765964.3811639}
\acmISBN{979-8-4007-2203-5/2026/08}

\usepackage[utf8]{inputenc} 
\usepackage{subcaption}
\usepackage{multirow}
\usepackage{colortbl}
\usepackage{color}
\usepackage{enumitem}
\usepackage{makecell}
\usepackage{graphicx} 

\begin{document}

\title[Understanding Students' Experiences with Natural Language Programming Tasks]{Understanding Student Perceptions, Mistakes, and Debugging Approaches when Solving Natural Language Programming Tasks}

\author{Victor-Alexandru P{\u a}durean}
\affiliation{%
  \institution{MPI-SWS}
  \city{Saarbr{\"u}cken}
  \country{Germany}
}
\email{vpadurea@mpi-sws.org}
\orcid{0009-0004-2998-096X}

\author{Kaitlin Riegel}
\affiliation{%
  \institution{University of Auckland}
  \city{Auckland}
  \country{New Zealand}
}
\email{kaitlin.riegel@auckland.ac.nz}
\orcid{0000-0002-8187-2016}

\author{Gweneth Barbre}
\affiliation{%
  \institution{Abilene Christian University}
  \city{Abilene}
  \state{TX}
  \country{USA}
}
\email{gab23c@acu.edu}
\orcid{0009-0007-3125-4168}

\author{Musa Blake}
\affiliation{%
  \institution{Abilene Christian University}
  \city{Abilene}
  \state{TX}
  \country{USA}
}
\email{mbb23c@acu.edu}
\orcid{0009-0001-2457-9147}

\author{Paul Denny}
\affiliation{%
  \institution{University of Auckland}
  \city{Auckland}
  \country{New Zealand}}
\email{paul@cs.auckland.ac.nz}
\orcid{0000-0002-5150-9806}

\author{Alkis Gotovos}
\affiliation{%
  \institution{MPI-SWS}
  \city{Saarbr{\"u}cken}
  \country{Germany}
}
\email{agkotovo@mpi-sws.org}
\orcid{0000-0002-3902-8890}

\author{Juho Leinonen}
\orcid{0000-0001-6829-9449}
\affiliation{%
  \institution{Aalto University}
  \city{Espoo}
  \country{Finland}
}
\email{juho.2.leinonen@aalto.fi}

\author{Stephen MacNeil}
\affiliation{%
  \institution{Temple University}
  \city{Philadelphia}
  \state{PA}
  \country{USA}
}
\email{stephen.macneil@temple.edu}
\orcid{0000-0003-2781-6619}

\author{James Prather}
\affiliation{%
  \institution{Abilene Christian University}
  \city{Abilene}
  \state{TX}
  \country{USA}
}
\email{james.prather@acu.edu}
\orcid{0000-0003-2807-6042}

\author{Adish Singla}
\affiliation{%
  \institution{MPI-SWS}
  \city{Saarbr{\"u}cken}
  \country{Germany}
}
\email{adishs@mpi-sws.org}
\orcid{0000-0001-9922-0668}

\renewcommand{\shortauthors}{Victor-Alexandru P{\u a}durean et al.}

\begin{abstract}
Learning to communicate with code-generating AI models is an emerging skill for novice programmers.  One recent pedagogical approach, Prompt Problems, has students solve computational tasks by writing natural-language prompts for code-generating AI models. However, little is known about the specific prompt-level mistakes novice programmers make, the kinds of computational details they fail to communicate, and what strategies they use to recover when generated code is incorrect. In a CS1 course, we studied attempts by more than 900 students to solve dialogue-based Prompt Problems. We analyzed student reflections, unsuccessful prompts, and reported debugging strategies. Compared to traditional coding tasks, students generally found prompting easier, more enjoyable, and better targeted at developing problem-solving skills. The most common mistakes are related to the omission of key details, suggesting both a failure to acknowledge their importance and over-reliance on AI to infer them. When prompts failed, students focused more on clarifying their intent and reflecting on the provided problem details than on tracing generated code or examining test cases.
\end{abstract}

\ccsdesc[500]{Social and professional topics~Computing education}
\ccsdesc[500]{Computing methodologies~Artificial intelligence}

\keywords{natural language programming, code-generating AI, Prompt Problems,  student perceptions}
\maketitle  


\section{Introduction}
\label{sec:intro}

The skills required by computing professionals are evolving as AI becomes increasingly integrated into the software development process \cite{ebert2023generative, santos2024impacts, song2024impactgenerativeaicollaborative}.  Generative AI (GenAI) is reshaping how software is created, tested, and maintained, and introducing new modes of interaction between humans and AI systems \cite{sauvola2024future}.  These include natural language communication between a developer and an AI programming assistant, such as GitHub Copilot.  Song et al. explored the impact of Copilot on open source projects, finding both project-level and individual productivity increases \cite{song2024impactgenerativeaicollaborative}. However, they noted part-time developers may struggle to generate good natural language prompts. There is growing recognition of the importance of effective prompting skills, with Ebert and Louridas claiming that finding the most appropriate way to frame a question to an AI assistant is becoming a ``new way of working'' for software practitioners \cite{ebert2023generative}.  Similarly, a recent ITiCSE working group led by Clear et al. interviewed $47$ IT professionals in New Zealand and Sweden and identified `prompt engineering' as a critical competency. 

\looseness-1The changing nature of software development with GenAI is also having an effect on how computing is taught from the introductory level \cite{denny2024cacm, liu2024teaching, fernandez2024cs1, amoozadeh2024student}. Where introductory programming (CS1) courses have traditionally focused on having students write correct code from clear problem statements, there is a renewed focus on core skills such as reading, evaluating, and interacting with AI-generated code \cite{prather2024beyond, feng2025redefining}. In their pioneering CS1-LLM work, Vadaparty et al. present a suite of new learning goals for an introductory programming course, which include prompt writing and engineering \cite{vadaparty2024cs1llm}. A case study in student-AI collaboration by Amoozadeh et al. revealed that the way students constructed their prompts had a considerable effect on success \cite{amoozadeh2024student}. Thus, crafting precise natural language instructions for AI models is becoming an essential skill, as unclear or incomplete prompts often lead to incorrect AI-generated solutions \cite{denny2023conversing}.  

\looseness-1A recently proposed approach for helping students develop prompt writing skills is Prompt Problems, where students are presented with a visual representation of a computational task for which they write a natural language prompt to generate solution code \cite{denny2024prompt}. Students reportedly find Prompt Problems engaging and beneficial for enhancing computational thinking skills \cite{prather2024breaking, kerslake2024integrating}. Such tasks allow students to focus on developing problem-solving skills, rather than spending cognitive resources writing syntactically correct code \cite{sweller1988cognitive}. Not only do these activities serve the broader purpose of aligning with industry changes, but they are pedagogically appropriate in targeting the development of specific and complex programming skills. However, prior implementations and investigations of student interactions with Prompt Problems have used tools that support a `zero-shot' approach, where each prompt attempt is required to contain all relevant information for the solution ~\cite{denny2024prompt, prather2024interactions}.  While suitable for very simple problems, modern AI coding assistants typically support dialogue-based interactions, where prior instructions form part of a larger context.  This is also a familiar interface for many students through experience with chat-based AI tools (e.g., ChatGPT).  

Computing education has a long history of focusing on both student perceptions and common mistakes when evaluating new pedagogical approaches. Thus, this paper explores how students perceive dialogue-based natural language programming tasks, particularly when compared to traditional coding activities.
Understanding students' experiences with Prompt Problems may help educators appropriately position natural language prompting within existing curricula. We further aim to delineate and catalog mistakes students make when constructing initial prompts.  While prior literature has identified common student mistakes in traditional programming tasks 
\cite{DBLP:conf/sigcse/KaczmarczykPEH10, qian2017students, lu2024identifying, rigby2020miss}, mistakes in crafting prompts for GenAI remain unexplored. Given the effectiveness of misconception-driven feedback (informed by mistakes) in traditional programming contexts \cite{gusukuma2018misconception}, understanding common mistakes could lead to targeted feedback strategies for prompt-based tasks. 
Finally, we note that prompt-based programming differs from traditional programming in how students must identify and debug errors. Traditional debugging often involves interpreting failed test cases and directly manipulating code, known to be challenging for novices \cite{yang2024decoding}. In contrast, debugging AI-generated outputs requires iterative refinement of prompts. Existing literature on debugging suggests that explicit instruction can improve novices' problem-solving capabilities in a traditional context \cite{hassan2024evaluating}. However, how students approach debugging when resolving errors that require refinement of natural language prompts is an open question. To address these gaps, we investigated students' experiences, mistakes, and reported debugging strategies when engaging with a tool presenting Prompt Problems and utilizing dialogue-based programming in an introductory C programming course. We were guided by the following questions:

\begin{enumerate}[label=\textbf{RQ\arabic*:},leftmargin=3em]
    \item How do students describe their experiences and perceived difficulty of solving dialogue-based Prompt Problems compared to traditional coding tasks?

    \item What types of mistakes do students make and how do these relate to each other when formulating natural language prompts for Prompt Problems?

    \item What strategies do students believe are most effective for refining dialogue-based prompts when the AI-generated code output is incorrect?
\end{enumerate}

Results were interpreted through the lens of Cognitive Load Theory \cite{sweller2011cognitive}. In this paper, we provide evidence that writing prompts allows students to focus more on the problem-solving process, and that they find it easier and more enjoyable compared to traditional code writing problems. We are able to outline the most common mistakes students make in their prompts, for example omitting the expected return type, and we report how these mistakes co-occur. Finally, we describe the strategies students use to correct errors when their initial prompt fails, with the most common being further clarifying their prompt. We discuss the use and challenges of more beneficial strategies such as reflecting on the problem, reviewing the test cases or error messages, and tracing the code. 



\section{Related Work}
\label{sec:related_work}

\subsection{Generative AI for Computing Education}
Recent advances in GenAI, particularly through code-generating large language models (LLMs), have significantly reshaped computing education~\cite{denny2024cacm,prather2023navigating}. These models have demonstrated their utility by identifying bugs~\cite{ macneil2024decoding}, repairing buggy programs~\cite{DBLP:journals/pacmpl/ZhangCGLPSV24}, generating clearer, student-friendly error messages and hints~\cite{DBLP:conf/sigcse/WangMP24,DBLP:conf/sigcse/0001HSRDPB23,DBLP:conf/edm/PhungCGKMSS23, brown2025howzat}, and providing detailed explanations of code concepts~\cite{DBLP:conf/icer/SarsaDH022,macneil23sigcse,leinonen2023comparing,bernstein2024like}. Research has also explored AI-supported debugging and pair programming, with GenAI tools used as intelligent assistants or programming partners, offering real-time guidance and immediate feedback~\cite{kazemitabaar2024codeaid,denny2023conversing,DBLP:conf/aied/MaWK23}.
As GenAI's importance in computing education rapidly grows, educators and researchers must assess new educational interventions for their impact on learning outcomes, ensuring they enhance rather than replace active student engagement and meaningful learning experiences~\cite{DBLP:conf/iticse/LiuYHBBL24,rawal2024hints,ahmed2025feasibility,DBLP:journals/corr/abs-2402-01580,prather2024widening}.

\subsection{Conversational Agents for Education}
\looseness-1GenAI-driven conversational agents are recognized as powerful educational tools. Rather than static hints or feedback, these systems engage learners in interactive dialogue, akin to peers and tutors. Agents based on generative models have been leveraged in tutoring systems that generate dialogues from lesson texts~\cite{DBLP:conf/aied/SchmuckerXAM24,acun2024gaienhanced}. These agents have also been used to simulate classroom interactions~\cite{DBLP:conf/lats/MarkelOLP23,lee2024generative} or model student behavior~\cite{DBLP:conf/edm/NguyenTS24}. For instance, one line of work uses generative agents to simulate students for teacher training~\cite{DBLP:conf/lats/MarkelOLP23,lee2024generative}, so pre-service teachers can practice responding to common questions and misconceptions in a low-risk setting. Another study showed how a GenAI-based agent framework can be used to facilitate language-learning games, boosting creativity and engagement through dialogue-based puzzles \cite{bailis2024wordplay}. These research directions highlight both the potential and challenges of such agents. They can increase student access to help and encourage more questions, yet ensuring the accuracy and pedagogical soundness of AI responses, while maintaining student trust, remains crucial \cite{khosravi2026building}. Despite these challenges, there is precedent for AI conversational agents to become valuable in education, offering personalized tutoring, timely feedback, and rich problem-solving dialogue at scale.

\subsection{EiPE and Prompt Problems}
Explain in Plain English (EiPE) tasks, where students receive a piece of code and then explain its behavior in their own words \cite{DBLP:conf/sigcse/MurphyMF12,whalley2006anaustralasian}, have long existed in computing education to assess and strengthen students' code comprehension. However, the inclusion of EiPE activities has commonly been limited, partly due to the difficulty of grading free-form explanations \cite{fowler2021autograding,DBLP:conf/icer/LiHFZZK22,DBLP:conf/aied/AzadCF0Z20}. Recently, LLMs have been explored in grading EiPE questions to support deploying these exercises at scale \cite{DBLP:conf/iticse/0001SFPB024}. Moreover, advances in GenAI have led to new kinds of programming exercises. Denny et al. introduced Prompt Problems, a novel activity where students solve programming tasks by writing natural language prompts for AI models rather than traditional code~\cite{denny2024prompt}. Their study, which was limited to `zero-shot' interactions, found that Prompt Problems positively engaged students, exposed them to new programming concepts, and promoted computational thinking without the distraction of syntax errors. 

Recent work explicitly links EiPE tasks with prompt-writing exercises. Smith et al. argued that explaining code and prompting for code are complementary learning activities, as both tasks require a deep understanding of what the code intends to achieve~\cite{smith2024prompting}. They suggested that combining EiPE-style questions with prompt-writing exercises can help students better grasp the behavior of AI, bridging the gap between reading, explaining, and generating code. In the same vein, Kerslake et al. introduced activities where students write prompts to generate code equivalent to given examples and found that these activities encourage students to engage a wider range of cognitive skills ~\cite{kerslake2024integrating}. In summary, both EiPE and Prompt Problems focus on developing students' understanding of program behavior through natural-language activities. By reducing barriers associated with coding syntax, these tasks can make programming more accessible to novices while supporting the development of other critical programming skills, such as code comprehension. Research is urgently required to determine whether these tasks are improving computing accessibility and supporting student comprehension as intended, and where problems exist, in order to develop meaningful learning opportunities.

\subsection{Mistakes in Programming and AI-Assisted Learning}
Mistakes are common in novice learners across any field, and new educational interventions always pose the risk of introducing new errors. Awareness of the types of mistakes students are likely to make allows educators to adjust their pedagogy and is therefore a critical area of investigation. In prompt-based programming tasks \cite{denny2024prompt,prather2024breaking}, students often struggle to articulate critical details in their prompts, resulting in incomplete or incorrect AI-generated solutions \cite{DBLP:journals/corr/abs-2501-10365, prather2024widening}, but this has yet to be comprehensively studied. Prior work has also shown that students often struggle with underspecified prompts~\cite{prather2024widening}, necessitating studying how they adjust their wording to guide AI-generated solutions~\cite{kazemitabaar2024codeaid}. Recent studies have further highlighted that novices frequently misinterpret or inadequately evaluate the code produced by LLMs, aggravating existing misconceptions or generating new misunderstandings about programming constructs and problem requirements~\cite{DBLP:conf/chi/NguyenBZGAF24, DBLP:journals/tochi/PratherRDBLLPFS24}. Misconceptions (i.e., incorrect mental models or flawed understandings about programming concepts) are a significant barrier for novice programmers~\cite{DBLP:conf/iticse/ChiodiniSGTSH21,DBLP:conf/sigcse/KaczmarczykPEH10} and may be informed by mistakes. Even though GenAI tools can help clarify misunderstandings by providing immediate corrections or tailored explanations~\cite{zvielgirshin2024thegood,macneil23sigcse, leinonen2023comparing}, they can also inadvertently introduce or reinforce misconceptions \cite{zvielgirshin2024thegood,prather2024widening}. Specifically, novices might accept AI-generated code or explanations without fully understanding their correctness, leading to new misconceptions and a false sense of competence~\cite{prather2024widening,DBLP:journals/corr/abs-2501-10365}.

Recently, Nam et al. detailed the mistakes developers make in prompting LLMs for code editing ~\cite{nam2025prompting}. However, this focused on the broader topic of everyday usage in the professional world, as opposed to simple programming problems, and we expect the types of mistakes novices make as they learn to prompt would be different from those of experts. Students' difficulties when refining their prompts, particularly in response to incorrect or misleading AI outputs, underscore the challenges associated with critical evaluation and iterative prompt revision~\cite{DBLP:conf/chi/NguyenBZGAF24}. Excessive reliance on AI output without careful checking mechanisms or guard-railing may reinforce faulty understandings. Consequently, examining students' strategies when refining prompts is an essential step in crafting exercises that foster deep conceptual learning in AI-assisted programming education. Investigating this will provide a foundation to further understand the scaffolding needed for these conversational agents to avoid reinforcing misconceptions. 

\subsection{Cognitive Load Theory}
Learning to program  can be particularly difficult for novices as they are presented with the challenge of simultaneously learning syntax, as well as abstract concepts and problem-solving skills. Cognitive Load Theory (CLT) can be applied to understanding how instructional design can facilitate learning through considering the limitations of student working memory \cite{sweller1988cognitive, sweller2011cognitive} (see ``New CLT'' ~\cite{duran2022cognitive}). Working memory is where new information is processed, and can then be transferred to long-term memory. However, working memory has only a small capacity. Heavy cognitive load (i.e., the effort exerted by the working memory) can hinder the processing and storage of new information in long-term memory, reducing learning. In computing education, CLT has been used to explain why novice programmers often experience cognitive overload when asked to comprehend or construct programs involving unfamiliar syntax, abstract logic, and problem-solving simultaneously. Mason et al. comment:
\begin{quote} A contributing factor to the difficulty of learning to program is that it inescapably combines aspects of logical concepts, algorithms, and programming language syntax [...] In having to deal simultaneously with all three and their interactions, programming novices are necessarily burdened with a very high level of load \cite[p.~45]{mason2016flipping}. \end{quote} 
Strategies aimed at promoting germane processing (i.e., schema construction) have been empirically validated through, for example, subgoal labeling \cite{morrison2016subgoals} and Parsons Problems \cite{denny2008evaluating, ericson2017solving, haynes2021problemsolving}. These reduce the superficial demands of a task to focus on schema-building, which, in turn, supports long-term learning \cite{duran2022cognitive}. 

Cognitive offloading describes the freeing up of working memory through the use of external tools or physical actions, such as calculators \cite{richmond2025benefits, risko2016cognitive}. Dialogue-based natural language prompting for code is an example of this, where students' intrinsic cognitive load is reduced by removing the task of producing correct syntax, instead allowing them to focus on problem-solving and take a higher-level view of the intended code. With this in mind, we would expect students to perceive these interactions as easier and more positively, given the freedom to focus on the larger problem, than they would in a traditional coding environment. However, students are more likely to cognitively offload (i.e., use the code produced by AI without understanding it), when they trust the tool with which they are working \cite{peng2025cognitive}. Thus, it is critical to evaluate whether this freeing up of cognitive resources is working as intended -- specifically, how do their problem-solving behaviors manifest, where do they perceive issues, and what errors arise in this new pedagogical environment? 



\section{Methodology}
\label{sec:method}

This section firstly describes the tool developed for conducting our study and explains how students were able to interact with it. We then provide information on the large-scale study context and design. Finally, we outline the survey questions, which solicited student reflections on the tasks, and the data analysis process. 

\subsection{Prompt Programming Tool}
\begin{figure*}[t!]
    \centering
    \begin{subfigure}{0.41\textwidth}
        \centering
        \begingroup
        \setlength{\fboxsep}{5pt}
        \setlength{\fboxrule}{0.4pt}
        \noindent\fcolorbox{gray!70!black}{purple!80!gray!5}{%
            \begin{minipage}{\dimexpr0.99\textwidth-2\fboxsep-2\fboxrule\relax}
                \small\looseness-1
                In this exercise, you will design a function foo that satisfies the given specification. You should begin by carefully looking at the provided specifications in the form of input-output pairs. Write your prompts to interact with the AI model and guide it to generate a correct program.
            \end{minipage}%
        }
        \par
        \endgroup
        \vspace{20.2mm}
        \includegraphics[width=\textwidth,trim={0mm -45mm 0mm 3mm},clip]{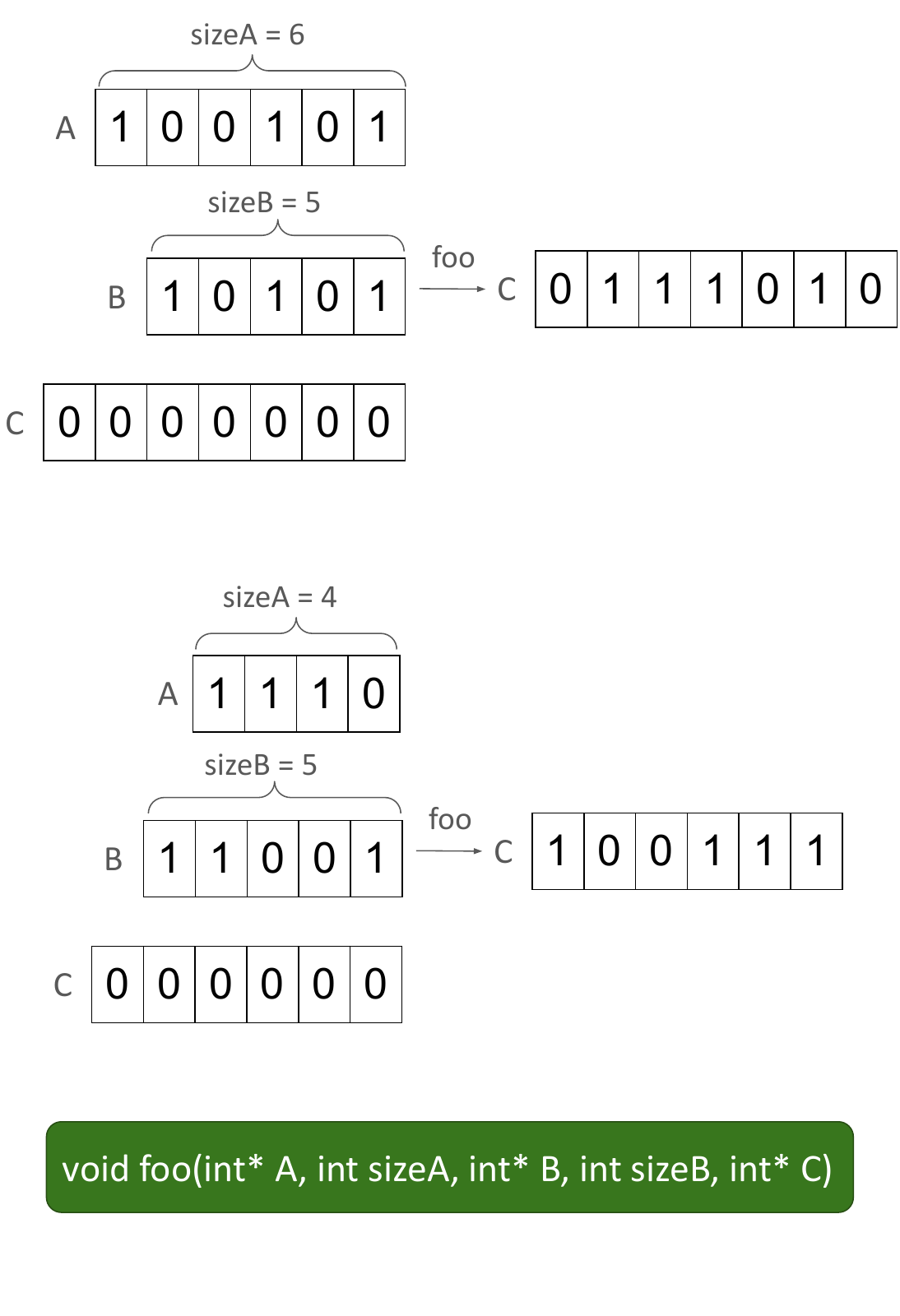}
        \caption{Problem description and input-output specs.}
        \label{fig.illustration-lab9.description}
    \end{subfigure}
    \hfill
    \begin{subfigure}{0.58 \textwidth}
        \centering
        \includegraphics[width=\textwidth]{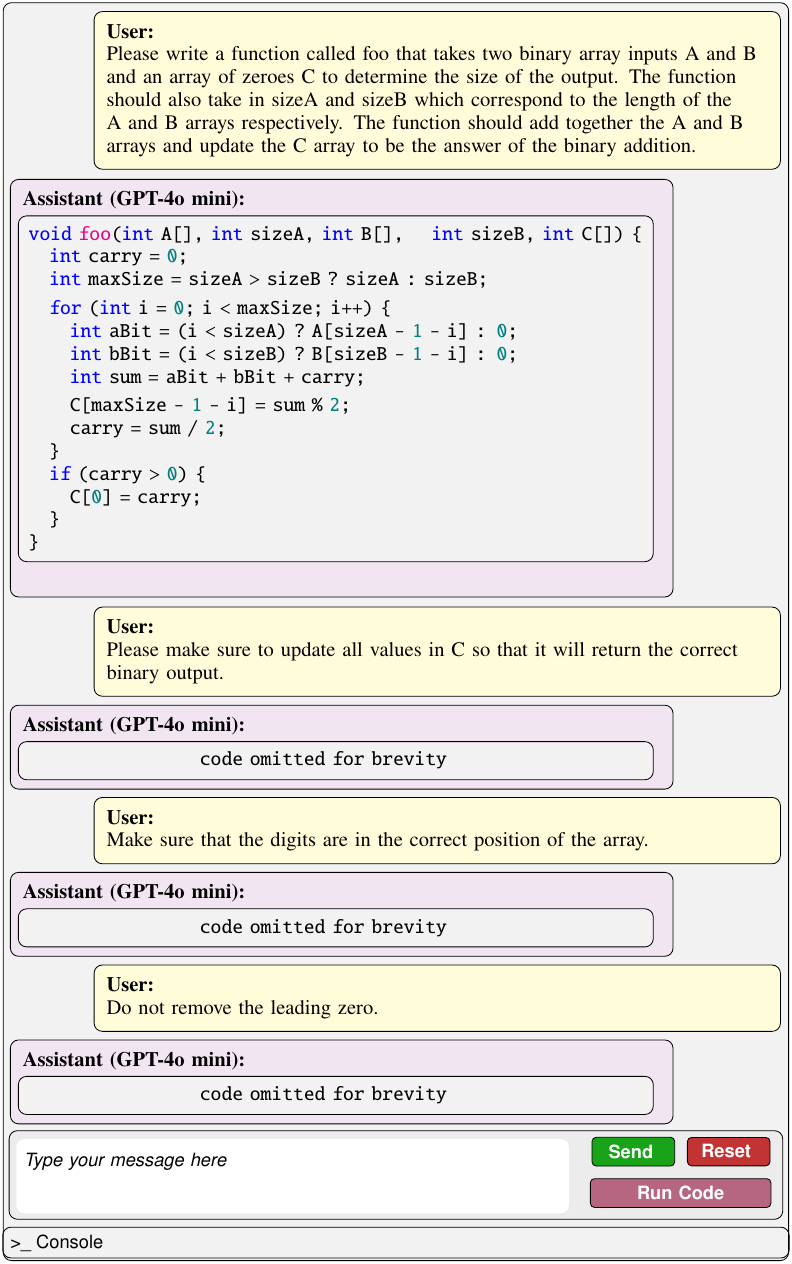}
        \caption{Example of a student's interaction through the chat interface.}
        \label{fig.illustration-lab9.interaction}
    \end{subfigure}
    \caption{Illustration of a student's iterative refinement process while successfully solving a Prompt Problem. \textbf{(a)} presents the problem description, including input-output specifications for the `binary addition' problem (B2-3). \textbf{(b)} showcases a student's interaction with the AI through a structured chat interface, highlighting how they refine their prompt step by step. We show the first AI-generated code output in full, which did not yet satisfy the problem specification mainly due to using an incorrect output size for array \texttt{C}. We omit later generated code for brevity. }
    \Description[Binary addition Prompt Problem and student refinement example]{Two-part figure illustrating a student's iterative refinement process for the binary addition Prompt Problem. Panel (a) shows the problem instructions and visual input-output specifications. The required function signature is void foo(int* A, int sizeA, int* B, int sizeB, int* C). The function takes two binary arrays A and B, their sizes sizeA and sizeB, and an output array C initialized with zeros, then updates C with the result of adding A and B as binary numbers. The first example shows A = [1, 0, 0, 1, 0, 1] with sizeA = 6, B = [1, 0, 1, 0, 1] with sizeB = 5, and C initially [0, 0, 0, 0, 0, 0, 0], producing C = [0, 1, 1, 1, 0, 1, 0]. The second example shows A = [1, 1, 1, 0] with sizeA = 4, B = [1, 1, 0, 0, 1] with sizeB = 5, and C initially [0, 0, 0, 0, 0, 0], producing C = [1, 0, 0, 1, 1, 1]. Panel (b) shows a chat-based interaction between a student and GPT-4o mini. The student first asks for a function named foo that adds binary arrays A and B and updates C. The first generated code is shown in full, but it does not satisfy the specification mainly because it uses the maximum of sizeA and sizeB as the output size instead of correctly updating all positions in C. The student then refines the prompt by asking the model to update all values in C, place the digits in the correct array positions, and preserve the leading zero. Later generated code is omitted for brevity.}
    \label{fig.illustration-lab9}
\end{figure*}

\begin{figure*}[tbp]
    \centering
    \begin{subfigure}{0.41\textwidth}
        \centering
        \begingroup
        \setlength{\fboxsep}{5pt}
        \setlength{\fboxrule}{0.4pt}
        \noindent\fcolorbox{gray!70!black}{purple!80!gray!5}{%
            \begin{minipage}{\dimexpr0.99\textwidth-2\fboxsep-2\fboxrule\relax}
                \small\looseness-1
                In this exercise, you will design a function foo that satisfies the given specification. You should begin by carefully looking at the provided specifications in the form of input-output pairs. Write your prompts to interact with the AI model and guide it to generate a correct program.
            \end{minipage}%
        }
        \par
        \endgroup
        \vspace{7.6mm}
        \includegraphics[width=\textwidth,trim={0mm 24mm 0mm 3mm},clip]{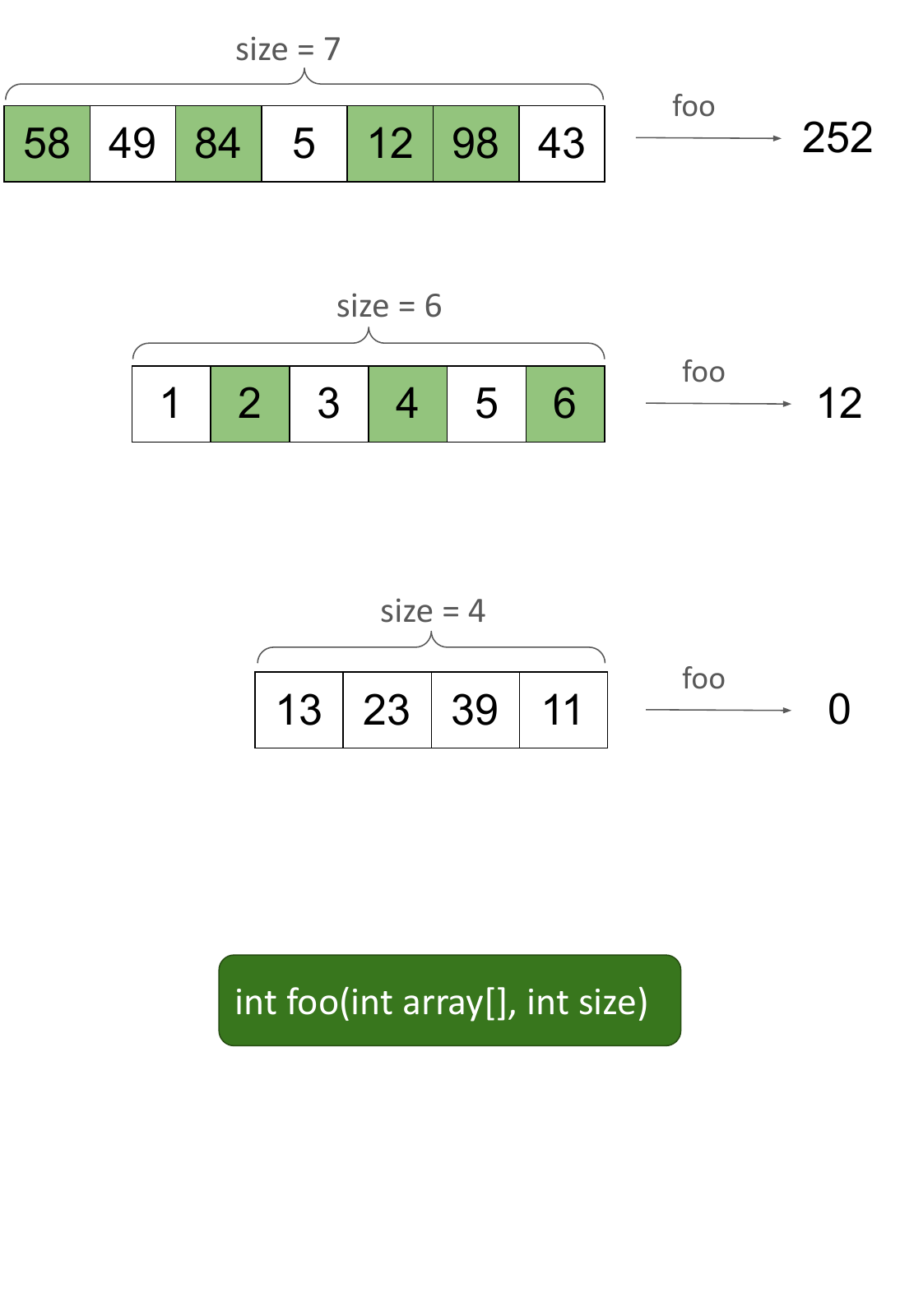}
        \caption{Problem description and input-output specs.}
        \vspace{2mm}
        \label{fig.illustration-lab7.description}
    \end{subfigure}
    \hfill
    \begin{subfigure}{0.58\textwidth}
        \centering
        \includegraphics[width=0.99\textwidth]{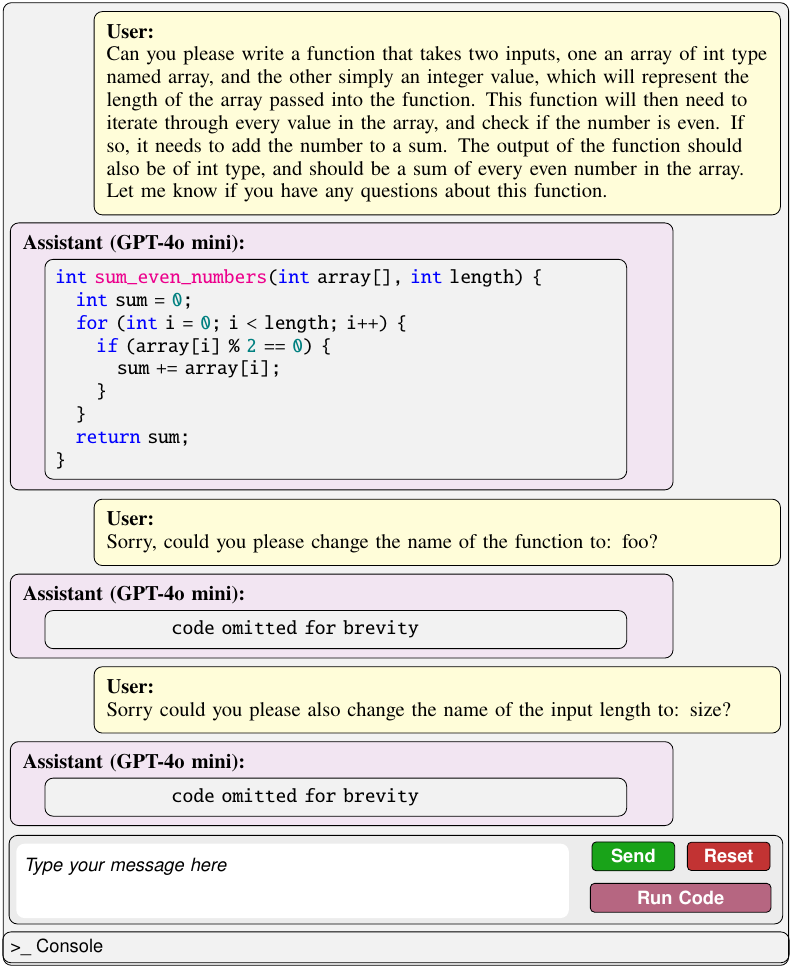}
        \caption{Example of a student's interaction through the chat interface.}
        \vspace{2mm}
        \label{fig.illustration-lab7.interaction}
    \end{subfigure}
    \caption{Illustration of a student's iterative refinement process while successfully solving a Prompt Problem. \textbf{(a)} presents the problem description, including input-output specifications for the `sum evens' problem (B1-2). \textbf{(b)} showcases a student's interaction with the AI through a structured chat interface, highlighting how they refine their prompt step by step. We show the first AI-generated code output in full, which did not yet satisfy the problem specification due to using an incorrect function name. We omit later generated code for brevity.}
    \Description[Sum evens Prompt Problem and student refinement example]{Two-part figure illustrating a student's iterative refinement process for the sum evens Prompt Problem. Panel (a) shows the problem instructions and visual input-output specifications. The required function signature is int foo(int array[], int size). The function takes an integer array and its size and returns the sum of all even numbers in the array. The first example shows array [1, 2, 3, 4, 5, 6] with size 6 returning 12. The second example shows array [58, 49, 84, 5, 12, 98, 43] with size 7 returning 252. The third example shows array [13, 23, 39, 11] with size 4 returning 0. Panel (b) shows a chat-based interaction between a student and GPT-4o mini. The student first asks for a function that takes an integer array named array and an integer representing the length of the array, iterates through the values, checks whether each value is even, and returns the sum of the even values as an integer. The first generated code is shown in full and implements the intended behavior, but it uses the function name sum_even_numbers and the parameter name length, which do not match the required signature. The student then refines the prompt by asking the model to change the function name to foo and the input length name to size. Later generated code is omitted for brevity.}
    \label{fig.illustration-lab7}
\end{figure*}

To conduct our analysis, we used Prompt Programming~\cite{DBLP:conf/iticse/Padurean0GS25}, a web-based tool for enabling students to solve Prompt Problems~\cite{denny2024prompt, prather2024breaking, DBLP:conf/iticse/Padurean0GS25} through dialogue-based interactions with a code-generating AI model. The tool enables an authentic experience through iterative, multi-turn interactions so students can refine prompts based on generated code, reflecting typical GenAI workflows~\cite{kazemitabaar2024codeaid, amoozadeh2024student, canvas}. We used GPT-4o~mini~\cite{GPT4omini} as the underlying AI model due to its accessibility and popularity. The interface displays each problem visually and provides a dialogue pane for prompts and AI-generated responses. The AI model did not receive the problem specification directly and it generated code only from the student's natural language messages and the preceding dialogue context. As a result, the assistant focused on writing or revising the code requested by the student, rather than providing pedagogical feedback or prompt-level feedback. In addition to a `Send' button, the pane offers a `Reset' button to restart the conversation and a `Run' button to execute the latest code against predefined tests (see Figures~\ref{fig.illustration-lab9}~and~\ref{fig.illustration-lab7}). The last generated block of code is always the one that is executed when requested, with an execution console showing the test results or any error.

\begin{figure*}[tbp!]
    \centering
    \begin{subfigure}{0.44\textwidth}
        \centering
        \includegraphics[width=\textwidth,trim={0mm 54mm 0mm 3mm},clip]{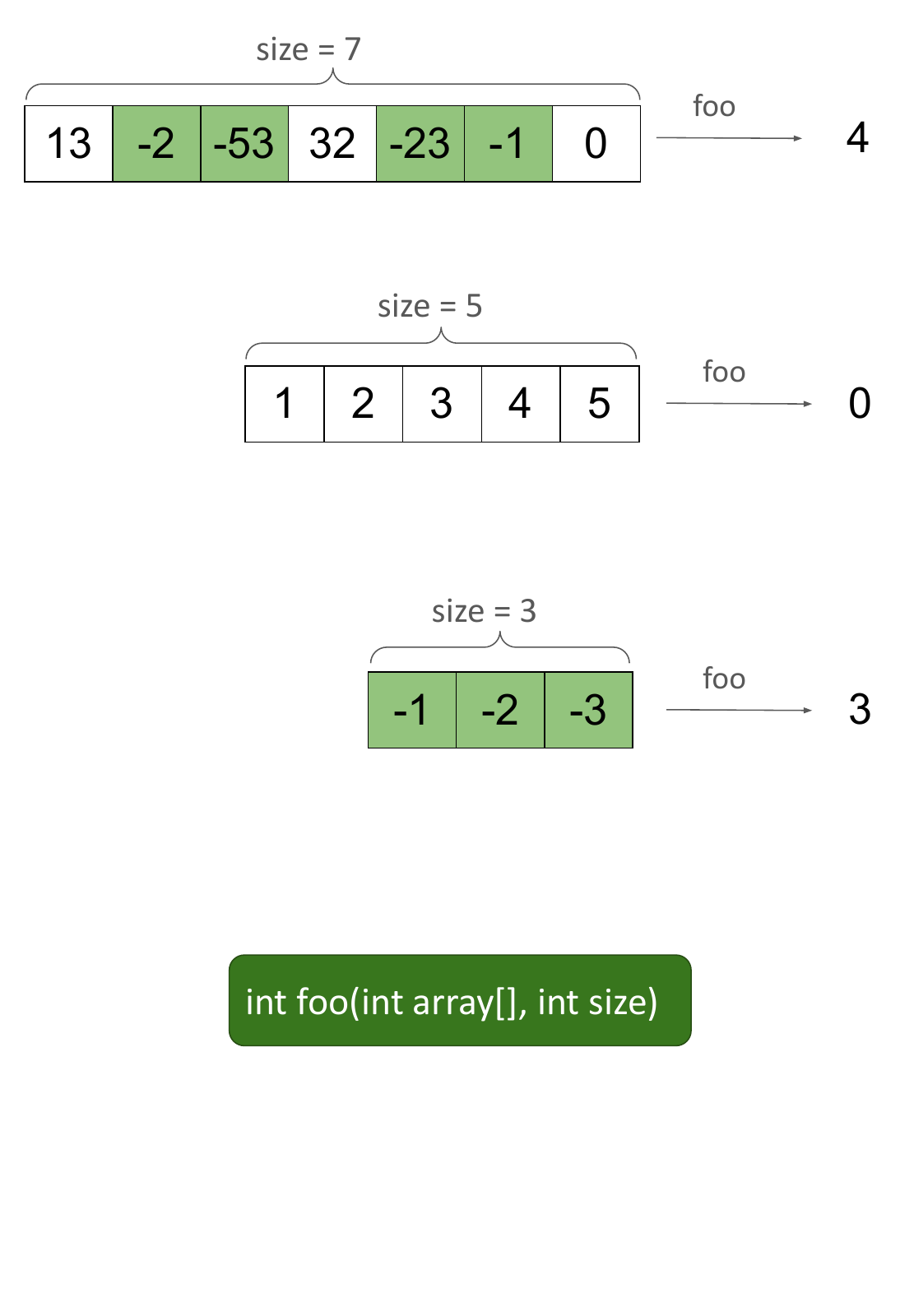}
        \vspace{-5mm}
        \caption{Specifications for `count negatives' problem (B1-1).}
        \label{fig.problems.count}
        \vspace{4.6mm}
    \end{subfigure}
    \hfill
    \begin{subfigure}{0.44\textwidth}
        \centering
        \includegraphics[width=\textwidth,trim={0mm 11mm 0mm 3mm},clip]{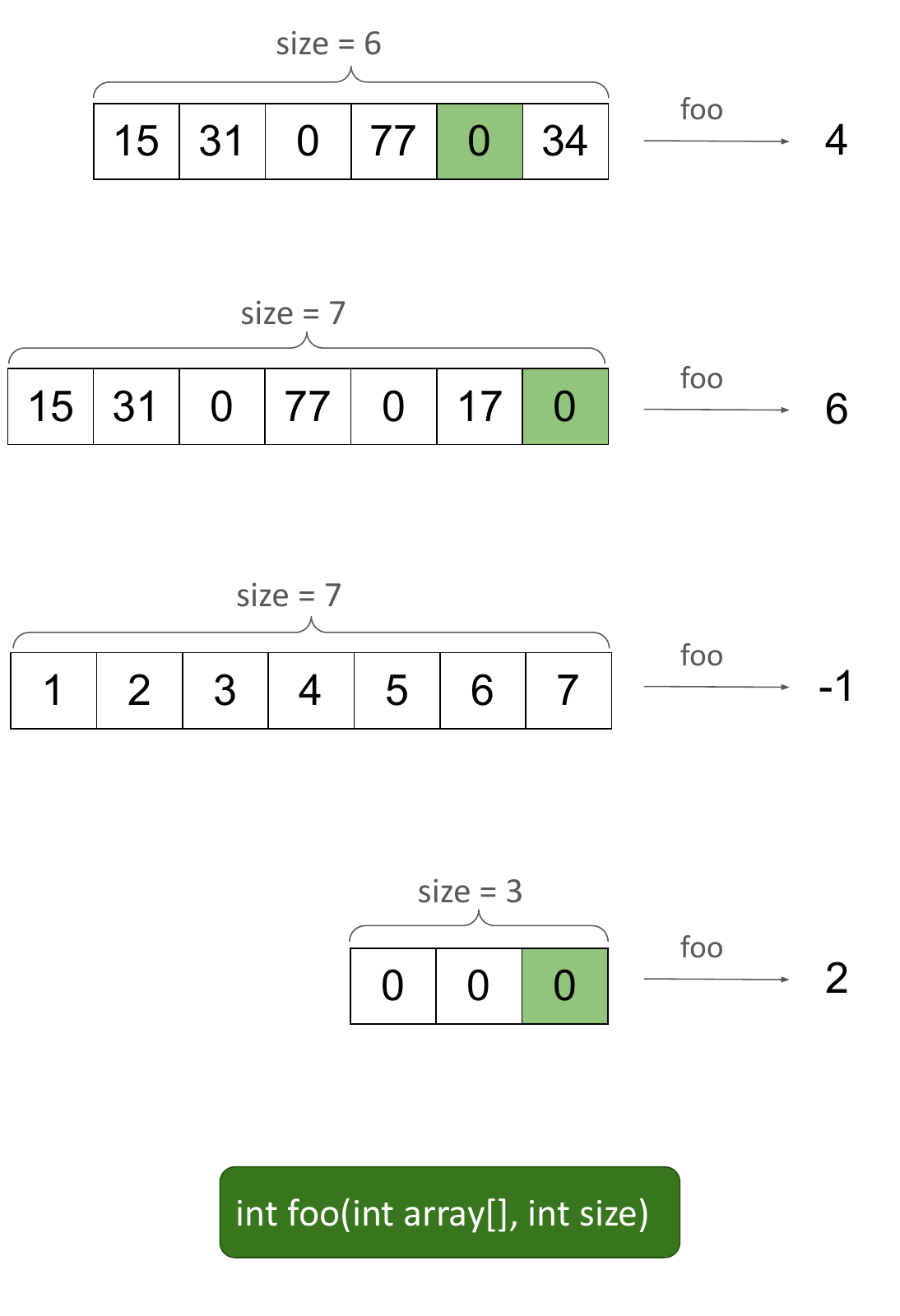}
        \vspace{-5mm}
        \caption{Specifications for `last zero's index' problem (B1-3).}
        \label{fig.problems.last}
        \vspace{4.6mm}
    \end{subfigure}
    \begin{subfigure}{0.44\textwidth}
        \centering
        \includegraphics[width=\textwidth,trim={0mm 45.5mm 0mm 3mm},clip]{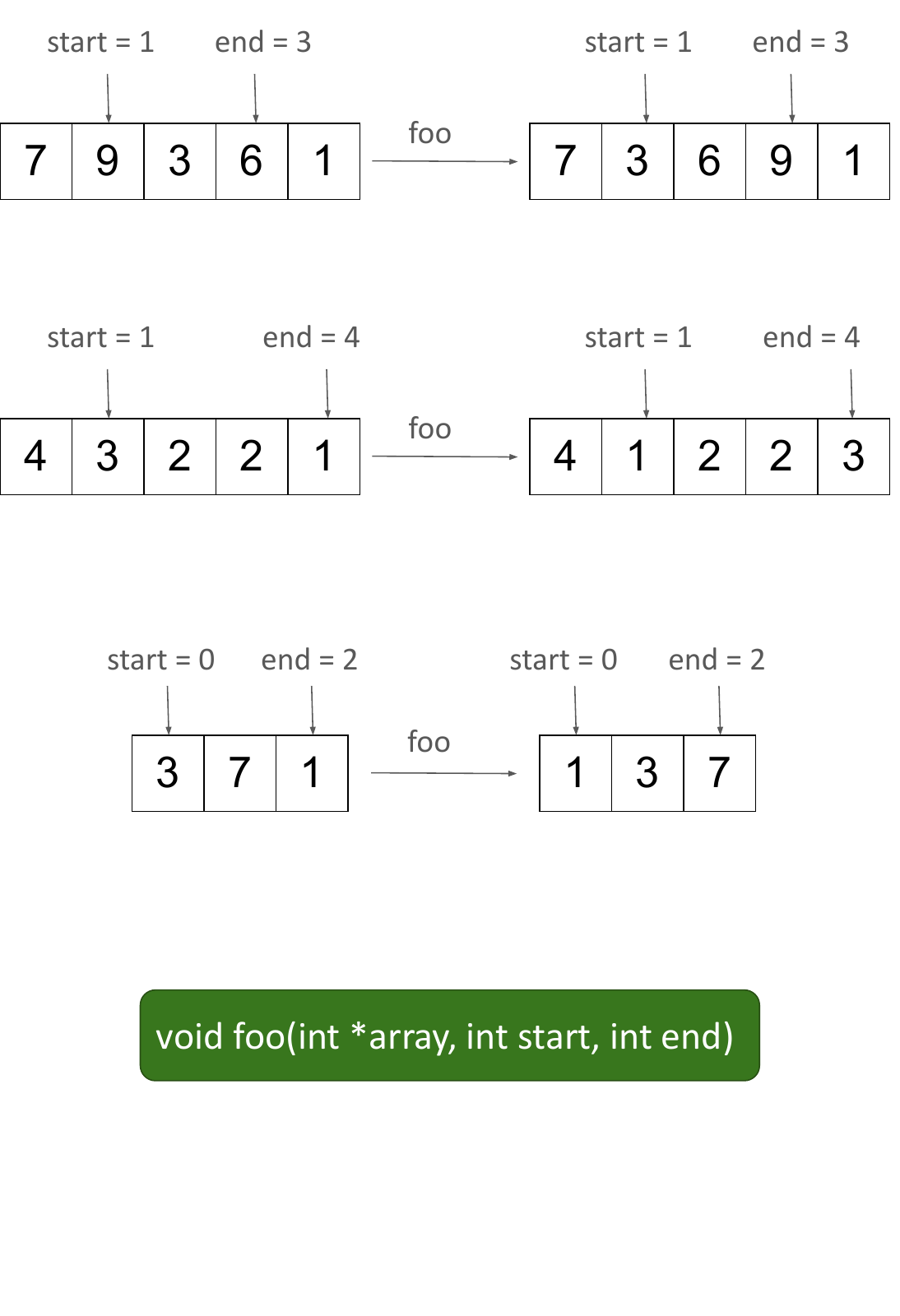}
        \vspace{-5mm}
        \caption{Specifications for `sort subarray' problem (B2-1).}
        \label{fig.problems.sort}
        \vspace{-2mm}
    \end{subfigure}
    \hfill
    \begin{subfigure}{0.44\textwidth}
        \centering
        \includegraphics[width=\textwidth,trim={0mm 42mm 0mm 3mm},clip]{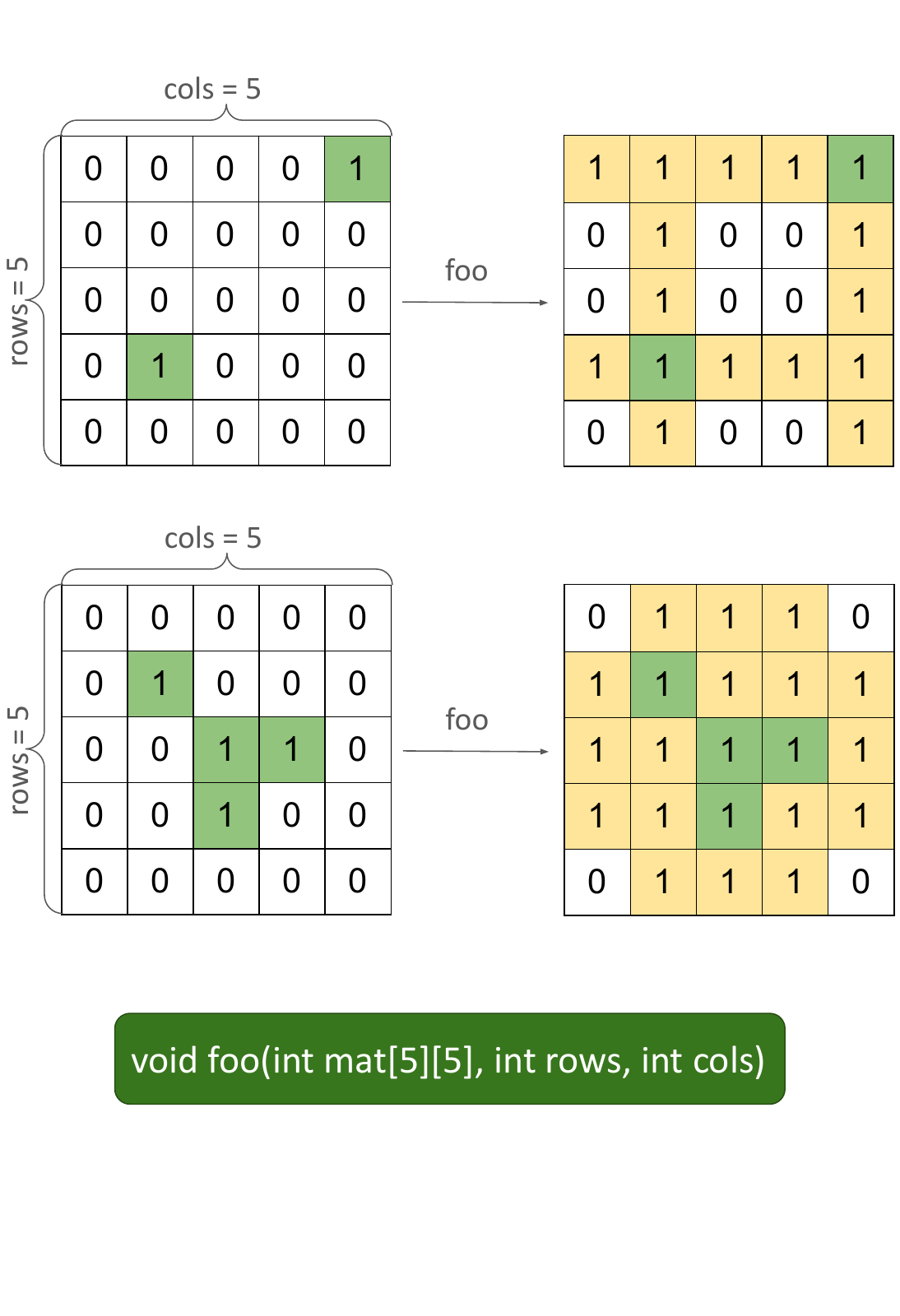}
        \vspace{-5mm}
        \caption{Specifications for `propagate ones' problem (B2-2).}
        \label{fig.problems.matrix}
        \vspace{-2mm}
    \end{subfigure}
    \caption{Illustration of four problems used in our study, with the remaining two shown in Figures~\ref{fig.illustration-lab9.description}~and~\ref{fig.illustration-lab7.description}. Problems in Figures~\ref{fig.problems.count},~\ref{fig.problems.last},~and~\ref{fig.illustration-lab7.description} comprised the first batch (B1), while problems in Figures~\ref{fig.problems.sort},~\ref{fig.problems.matrix},~and~\ref{fig.illustration-lab9.description} comprised the second batch (B2).}
    \Description[Four Prompt Problem specifications used in the study]{Four-part figure showing visual input-output specifications for four Prompt Problems used in the study. Panel (a) shows the count negatives problem, with function signature int foo(int array[], int size). The function takes an integer array and its size and returns the number of negative values. The examples shown are array [13, -2, -53, 32, -23, -1, 0] with size 7 returning 4; array [1, 2, 3, 4, 5] with size 5 returning 0; and array [-1, -2, -3] with size 3 returning 3. Panel (b) shows the last zero's index problem, with function signature int foo(int array[], int size). The function takes an integer array and its size and returns the index of the last zero, or -1 when no zero is present. The examples shown are [15, 31, 0, 77, 0, 17, 0] with size 7 returning 6; [15, 31, 0, 77, 0, 34] with size 6 returning 4; [1, 2, 3, 4, 5, 6, 7] with size 7 returning -1; and [0, 0, 0] with size 3 returning 2. Panel (c) shows the sort subarray problem, with function signature void foo(int *array, int start, int end). The function modifies the array by sorting only the elements from index start through index end, while leaving the remaining elements unchanged. The examples shown are [7, 9, 3, 6, 1] with start 1 and end 3 becoming [7, 3, 6, 9, 1]; [3, 7, 1] with start 0 and end 2 becoming [1, 3, 7]; and [4, 3, 2, 2, 1] with start 1 and end 4 becoming [4, 1, 2, 2, 3]. Panel (d) shows the propagate ones matrix problem, with function signature void foo(int mat[5][5], int rows, int cols). The function modifies a matrix so that every row and column containing at least one 1 is filled with 1s at the corresponding positions. The first example has a 5 by 5 matrix with 1s in the top-right cell and in row 4, column 2, producing an output where the first and fourth rows are all 1s and the second and fifth columns are filled with 1s. The second example has 1s at row 2, column 2; row 3, columns 3 and 4; and row 4, column 3, producing an output where rows 2, 3, and 4 are all 1s, and columns 2, 3, and 4 are filled with 1s. The first batch of problems includes count negatives, sum evens, and last zero index, while the second batch includes sort subarray, propagate ones, and binary addition.}
    \label{fig.problems}
\end{figure*}

\looseness-1To examine how students refine prompts across different levels of complexity, we used six Prompt Problems inspired by standard introductory problems and recent literature~\cite{denny2024prompt}. The first three problems involved determining the count of negative numbers in an array (Figure~\ref{fig.problems.count}), computing the sum of all even elements (Figure~\ref{fig.illustration-lab7.description}), and identifying the position of the last zero in an array (Figure~\ref{fig.problems.last}). These tasks focused on fundamental array traversal and filtering operations, where the goal was to produce a single integer result based on specific conditions. The remaining problems required working with more structured data: sorting a specified section of an array while leaving the rest unchanged (Figure~\ref{fig.problems.sort}), modifying a matrix so that every row and column containing at least one 1 is entirely filled with 1s (Figure~\ref{fig.problems.matrix}), and performing binary addition using arrays to represent numbers (Figure~\ref{fig.illustration-lab9.description}). These tasks emphasized indexing precision, conditionally modifying data, and handling in-place updates, offering a valuable setting for studying how students refine prompts when dealing with additional constraints.

During the study, students solved each prompt-based task by interacting with the tool through natural language prompts. As illustrated in Figures~\ref{fig.illustration-lab9} and~\ref{fig.illustration-lab7}, students first viewed a visual depiction of a computational task and then wrote an initial prompt explaining the desired behavior. After reviewing the generated code produced by GPT-4o~mini, students could either test it immediately using on-demand code execution or engage in iterative dialogue by refining their prompts in subsequent messages. However, the version of Prompt Programming we used did not allow students to edit the code directly. If students felt their interaction had reached a problematic state, they could reset the dialogue and start again. Students were not told how to construct prompts or given anything beyond basic instructions on interacting with the tool. 

\subsection{Study Design}
We deployed our study during Semester 2, 2024 in an introductory C programming class designed for engineering students at the University of Auckland, with 1,031 enrolled students. Two batches of problems and corresponding reflection questions were integrated into two separate, week-long laboratory sessions during the seventh and ninth weeks of semester. Analysis of the data for this study was approved by the university ethics committee (approval number UAHPEC25279). To systematically collect data addressing our research questions, we grouped the six problems into two batches. In the first batch (B1), students solved three simpler array-based Prompt Problems (B1-1, B1-2, and B1-3, for brevity). Immediately after completing these, students answered three reflection questions (Ref1, Ref2, Ref3) designed to capture their experiences solving Prompt Problems compared to traditional coding tasks, and their perceived difficulty in guiding the AI model.

\begin{enumerate}[label=\textbf{Ref\arabic*:},leftmargin=*]
    \item Please comment on your experience solving these tasks compared to traditional programming tasks.
    \item Please comment on how easy or difficult it was to guide the AI model to generate the desired code.
    \item It was easier to solve these problems by writing natural language than by writing code. \emph{(Five-point Likert item)}.
\end{enumerate}

In the second batch (B2), students approached the other three more complex problems (B2-1, B2-2, and B2-3, for brevity) and were required to solve at least one of the three. After their attempt at these more challenging problems, students had the opportunity to answer an additional reflection question (Ref4) specifically designed to capture their debugging strategies.

\begin{enumerate}[label=\textbf{Ref\arabic*:},leftmargin=*,start=4]
    \item Please describe your strategies when the AI-generated output was incorrect. What information did you focus on most when refining your prompts?
\end{enumerate}

We logged information on student success rate, messages exchanged with the tool, code executions, and usage of the reset button (starting a new conversation). This study design allowed us to examine how students described their experience with and perceived difficulty of prompt-based tasks, particularly compared to traditional coding (RQ1). We also collected data on what common mistakes occurred through analysis of students' initial unsuccessful prompts (RQ2), and their perspectives on effective approaches to debugging and refinement of incorrect AI-generated code (RQ3). 

Although reflection questions pertaining to the lab questions were optional, all students were required to solve all three problems from the first batch (B1), and at least one from the second batch (B2) in order to receive marks contributing approximately 1\% towards their course grade.

\begin{table*}[t!]
    \caption{Descriptive statistics for students who successfully completed the requirements for B1 (\textit{N} = 887) and B2 (\textit{N} = 876).}
    \label{fig.table_1}
\begin{tabular}{llcccc}
\hline
{\color[HTML]{000000} }                     & {\color[HTML]{000000} }                & {\color[HTML]{000000} Mean (SD)}   & {\color[HTML]{000000} Range}       & {\color[HTML]{000000} Skewness (SE)} & {\color[HTML]{000000} Kurtosis (SE)} \\ \cline{3-6} 
{\color[HTML]{000000} }                     & {\color[HTML]{000000} Conversations}   & {\color[HTML]{000000} 1.35 (0.61)} & {\color[HTML]{000000} 1.00, \phantom{0}6.33}  & {\color[HTML]{000000} 2.93 (.08)}    & {\color[HTML]{000000} 12.11 (.16)}   \\
{\color[HTML]{000000} }                     & {\color[HTML]{000000} Messages}        & {\color[HTML]{000000} 4.96 (3.79)} & {\color[HTML]{000000} 2.00, 34.00} & {\color[HTML]{000000} 2.70 (.08)}    & {\color[HTML]{000000} 10.08 (.16)}   \\
\multirow{-3}{*}{{\color[HTML]{000000} B1}} & {\color[HTML]{000000} Code executions} & {\color[HTML]{000000} 1.80 (1.25)} & {\color[HTML]{000000} 1.00, 12.33} & {\color[HTML]{000000} 3.14 (.08)}    & {\color[HTML]{000000} 13.94 (.16)}   \\ \hline
{\color[HTML]{000000} }                     & {\color[HTML]{000000} Conversations}   & {\color[HTML]{000000} 1.28 (1.12)} & {\color[HTML]{000000} 0.33, 11.00} & {\color[HTML]{000000} 3.27 (.08)}    & {\color[HTML]{000000} 17.68 (.17)}   \\
{\color[HTML]{000000} }                     & {\color[HTML]{000000} Messages}        & {\color[HTML]{000000} 6.63 (6.51)} & {\color[HTML]{000000} 0.67, 58.00} & {\color[HTML]{000000} 2.55 (.08)}    & {\color[HTML]{000000} 10.25 (.17)}   \\
\multirow{-3}{*}{{\color[HTML]{000000} B2}} & {\color[HTML]{000000} Code executions} & {\color[HTML]{000000} 2.77 (2.69)} & {\color[HTML]{000000} 0.33, 27.33} & {\color[HTML]{000000} 2.59 (.08)}    & {\color[HTML]{000000} 11.59 (.17)}   \\ \hline
\end{tabular}
\end{table*}

\subsection{Data Analysis}

Prompting interactions outlined above were collected following the labs' submission deadline (students who submitted late were excluded). A very small fraction of data -- nine conversations between five students across the entire dataset -- was removed for being problematic (i.e., networking issues that disrupted a student-LLM conversation). The labs were designed so students could complete the tasks in person or at home in their own time. Therefore, time spent on the problem was not necessarily a meaningful measure and was not used for filtering or analysis in this study. 

\looseness-1A codebook approach to inductive thematic analysis was employed for the three survey items soliciting qualitative data (Ref1, Ref2, Ref4) \cite{braun2022conceptual, clarke2014thematic}. An experienced researcher led a team of three, who worked together to inductively code student responses. When no new codes appeared, the data was considered thematically saturated \cite{rahimi2024saturation, saunders2018saturation}. All codes were updated and re-evaluated on each response to ensure complete and consistent coverage. As the three researchers did this, they discussed any disagreements and decided on outcomes together. Reliability was established through continuous discussion and consensus-building, consistent with established practices in qualitative research~\cite{mcdonald2019reliability}. Saturation was reached at or before response $160$ for all three reflection questions, so the researchers chose to continue to $200$ total coded responses, during which time no additional codes were identified. Responses to Ref4 from students who solved the problems immediately were not included in the final analysis, as it asked them to reflect on their prompt-refinement strategies. After coding $200$ responses for Ref1, Ref2, and Ref4, the codes were consolidated into themes. Codes for Ref1 ($21$) and Ref2 ($17$) were grouped together into $14$ themes, since both reflection questions were about user experience with the tool and contribute to RQ1. Codes for Ref4 ($25$) were consolidated into $11$ separate themes to address RQ3.

In order to identify mistakes made by students (RQ2), natural language prompts written by students were coded. Three researchers worked to create a list of codes representing 
required elements that should be present in every successful prompt (with a single category for extraneous elements that should be absent). We did not begin with previous literature, since mistakes made while prompting are still poorly understood -- prior studies have focused on self-reported difficulties \cite{kerslake2024integrating} and quantified prompt modifications (e.g., number of words or characters changed) \cite{DBLP:conf/chi/NguyenBZGAF24, prather2024interactions}, but no work has systematically examined the content of prompts or their errors. Researchers began by referring to the Prompt Problems and making a list of possible mistakes as initial codes, while further mistakes emerged inductively. Prompts were re-coded as new codes were added. Codes were added until reaching saturation and consolidated into a final set of codes. In order to get a broad representation of student mistakes, half of all incorrect first prompts for each problem were coded, for a total of 1,286 prompts. Each prompt was coded with a particular code if it was \textit{missing} that element. 



\section{Results}
\label{sec:results}

\subsection{Descriptive Statistics}

\looseness-1We analyzed the two problem batches to obtain an overview of student engagement. In B1, $920$ students attempted at least one problem, with $887$ succeeding on all three (required to obtain marks). In B2, $916$ students attempted at least one problem with $876$ succeeding on at least one (required to obtain marks). Table~\ref{fig.table_1} presents the descriptive statistics for students who successfully completed the exercise, including the average number of messages, conversations, and code executions per problem. We perceived this as more meaningful than presenting data on all students who attempted the problems, since this could bring significant noise (i.e., incomplete tasks after varying degrees of effort). However, we note that in B2, some students stopped after correctly answering one problem and meeting the requirements (problems solved in B2: \textit{M} = 1.55, \textit{SD} = 0.79). Overall, the results align with B2 problems requiring more effort, with greater variance in the variables (though, notably, there is a large variation in the number of messages sent for both batches). 

\subsection{RQ1: Comparison to Traditional Coding and Difficulty}\label{sec:results.rq1}

\begin{figure*}[t!]
\centering
	\includegraphics[width=0.9\linewidth]{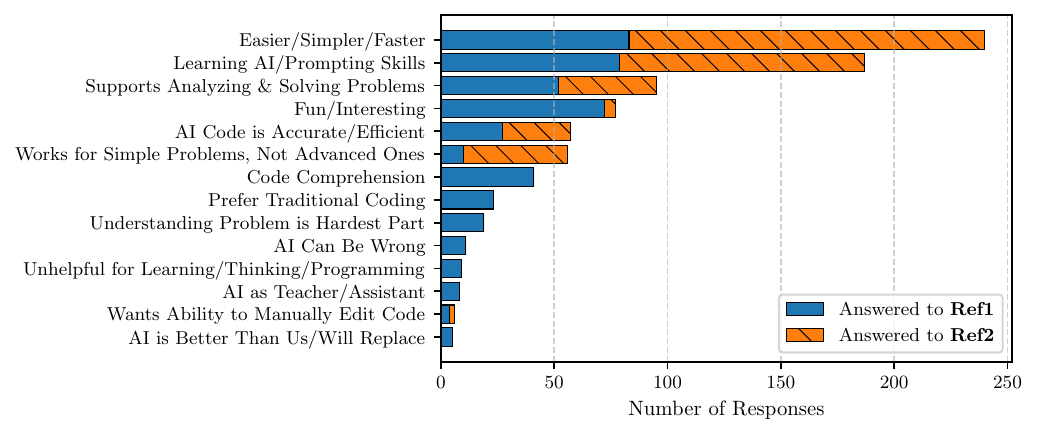}
     \vspace{-4mm}
     \caption{Themes identified in students' responses to Ref1 and Ref2. Bars show the number of coded occurrences for each theme across 200 student reflections, with stacked segments indicating whether the theme appeared in responses to Ref1 or Ref2.
     }
    \Description[RQ1 thematic analysis of student reflections on prompt-based tasks]{Horizontal stacked bar chart showing themes from 200 tagged student reflections about their experiences with Prompt Problems and the difficulty of guiding the AI model. The x-axis shows the number of responses, ranging from 0 to 250, and the y-axis lists fourteen themes. Each horizontal bar is split by reflection source, distinguishing responses to Ref1, about experience compared to traditional programming tasks, and Ref2, about how easy or difficult it was to guide the AI model. The themes, ordered from most to least frequent, are Easier/Simpler/Faster; Learning AI/Prompting Skills; Supports Analyzing and Solving Problems; Fun/Interesting; AI Code is Accurate/Efficient; Works for Simple Problems, Not Advanced Ones; Code Comprehension; Prefer Traditional Coding; Understanding Problem is Hardest Part; AI Can Be Wrong; Unhelpful for Learning/Thinking/Programming; AI as Teacher/Assistant; Wants Ability to Manually Edit Code; and AI is Better Than Us/Will Replace. The largest bars indicate that students most often described the tasks as easier, simpler, or faster than traditional coding, as helping them learn AI or prompting skills, as supporting problem analysis and problem solving, and as fun or interesting. Smaller bars capture concerns or reservations, including preference for traditional coding, difficulty understanding the problem, the possibility that AI can be wrong, doubts about learning value, desire for manual code editing, and concerns about AI replacing programmers.}
    \label{fig.rq1}
\end{figure*}

Figure~\ref{fig.rq1} presents the final themes for students' diverse reflections on their experiences with writing prompts while solving Prompt Problems. The bars are grouped by theme, and within each bar the counts are stacked to show the separate contributions of Ref1 and Ref2. Numerous positive perceptions were reported from the students. The most frequently reported advantage was that it was \textbf{Easier/Simpler/Faster}, that is, using natural language made writing and modifying their code easier than traditional code writing. For example, one student wrote, ``It's super easy for me personally, like just by entering what will be your input, and what [is] expected to be your output, the correct code will be generated.'' Many also found the tasks \textbf{Fun/Interesting}. Significantly, students suggested the experience \textbf{Supports Analyzing \& Solving Problems}, capturing their reflections around how the task eschewed difficult syntax in favor of natural language descriptions, which allowed students to concentrate more on high-level problem solving. It also included student sentiment that they were learning and solving the problem through attempting it. One student wrote, ``As a relatively inexperienced coder, I understand the logic behind the tasks but often struggle with the precise syntax. Therefore I found using AI to be extremely beneficial." Further, students noted the tool's ability to support and scaffold \textbf{Code Comprehension} by taking natural language prompts and generating code, thus connecting the two for learners. 

Outside of traditional computing education skills, students felt they were \textbf{Learning AI/Prompting Skills}, reflected in comments on how the task taught them to use AI or write prompts, how to write robust and specific prompts, and how to articulate difficult concepts in natural language. One student commented, 
\begin{quote}
Rather than just reading a problem that we are meant to solve we had to figure out the problem ourselves and the easy part was getting the AI to fulfill our prompt \ldots{} it builds critical thinking for coding which I believe is important for more real-world situations as in real-world situations we won't always be handed the issue but rather given the task of figuring out the issue ourselves and being able to solve it.\end{quote} 

Many students reflected positively on the accuracy of AI code, its efficiency, that it generates more efficient code than they would write, and that it often infers details missing from the prompt but were ultimately necessary (\textbf{AI Code is Accurate/Efficient}). However, a subset of students expressed concern or reluctance. \textbf{AI Can Be Wrong} refers to comments about how LLMs can generate incorrect code, which can be difficult for novices to detect. \textbf{Works for Simple Problems, Not Advanced Ones} refers to the sentiment that the tool works well for easy problems but not quite well enough for larger or more advanced problems, such as those requiring multiple functions. Some students also felt that the complexity depended on the student's ability to find the pattern or understand the problem in the first place. While acknowledging the accuracy and effectiveness of AI-generated solutions for simple tasks, several students highlighted growing uncertainty and reduced trust as problems increased in complexity, captured in the following reflection: ``These functions were so simple, but I did not need to provide too much context, and it managed to fill in the gaps \ldots{} however, I would tend to trust it less as the complexity increases.''

Despite recognizing the convenience of prompt-based programming, there were students who indicated they \textbf{Prefer Traditional Coding}. One commented, ``So cool and epic, but I prefer writing code even if prompt programming is easier.'' Some suggested a combined prompting approach (\textbf{Wants Ability to Manually Edit Code}), expressing frustrations and wanting the ability to manually edit the AI-generated code as opposed to prompt-refinement. There were several smaller themes. Some students noted that \textbf{Understanding Problem is Hardest Part}, highlighted in reflections that determining \textit{what} to solve was more difficult than \textit{how} to solve. In direct conflict, others argued that these tasks were \textbf{Unhelpful for Learning/Thinking/Programming}. Finally, a small number of students discussed the role of AI, with some identifying \textbf{AI as Teacher/Assistant} and others commenting about the future and their place in it, specifically how they believe AI is going to replace human programmers (\textbf{AI is Better Than Us/Will Replace}).

\begin{table*}[t!]
    \caption{Per problem means (standard deviations) for the number of conversations, total messages sent over all conversations, success rate, and number of mistakes present in the initial prompt for all students who submitted an initial prompt resulting in incorrect code.}
    \label{fig.table_2}
\begin{tabular}{lcccccc}
\hline
\textbf{}      & B1-1        & B1-2        & B1-3          & B2-1          & B2-2         & B2-3          \\ \cline{2-7} 
Conversations  & 1.51 (1.07) & 1.61 (1.14) & \phantom{0}2.20 (\phantom{0}1.86)   & \phantom{0}2.26 (\phantom{0}2.53)   & \phantom{0}2.01 (1.80)  & \phantom{0}2.51 (\phantom{0}3.03)   \\
Total Messages & 6.98 (7.16) & 7.01 (6.86) & 11.90 (11.39) & 12.07 (12.70) & 10.34 (9.92) & 15.85 (18.76) \\
Success rate & 0.98 (0.16) & 1.00 (0.00) & \phantom{0}0.98 (\phantom{0}0.14)   & \phantom{0}0.75 (\phantom{0}0.44)   & \phantom{0}0.60 (0.49)  & \phantom{0}0.44 (\phantom{0}0.50)   \\
Total Mistakes       & 4.81 (1.65) & 5.24 (1.72) & \phantom{0}5.18 (\phantom{0}1.52)   & \phantom{0}3.84 (\phantom{0}1.52)   & \phantom{0}3.27 (1.46)  & \phantom{0}3.28 (\phantom{0}1.51)   \\ \cline{2-7} 
\textit{n} =          & 162         & 180         & 243           & 329           & 194          & 163           \\ \hline
\end{tabular}
\end{table*}

\begin{table*}[t!]
    \caption{Frequency ($\%$) of missing prompt elements from the first submitted incorrect prompt (i.e. not leading to correct code) by problem in B1 and B2. Color indicates the top three missing prompt elements by problem.}
    \label{fig.table_3}
\begin{tabular}{lllllll}
\hline
                                      & \multicolumn{6}{c}{\% Per Problem}                                                                                                                                                            \\ \cline{2-7} 
                                      & \multicolumn{1}{c}{B1-1}      & \multicolumn{1}{c}{B1-2}      & \multicolumn{1}{c}{B1-3}      & \multicolumn{1}{c}{B2-1}      & \multicolumn{1}{c}{B2-2}      & \multicolumn{1}{c}{B2-3}      \\ \cline{2-7} 
Appropriate expected output/return          & 64.20                         & \cellcolor[HTML]{94DCF8}83.33 & \cellcolor[HTML]{61CBF3}89.30 & NA                            & NA                            & NA                         \\
Appropriate functionality explanation & 43.83                         & 73.33                         & 79.84                         & \cellcolor[HTML]{94DCF8}93.31 & 33.51                         & \cellcolor[HTML]{CAEDFB}57.67 \\
Appropriate required input(s)         & \cellcolor[HTML]{CAEDFB}79.01 & 73.89                         & 69.55                         & 40.73                         & \cellcolor[HTML]{CAEDFB}52.06 & 28.83                         \\
Correct order of arguments            & 50.62                         & 40.00                         & 38.68                         & 25.23                         & 29.90                         & \cellcolor[HTML]{94DCF8}58.28 \\
Specified argument names              & \cellcolor[HTML]{94DCF8}89.51 & \cellcolor[HTML]{CAEDFB}81.67 & \cellcolor[HTML]{CAEDFB}82.30 & 35.56                         & \cellcolor[HTML]{94DCF8}80.41 & 57.06                         \\
Specified function return type        & \cellcolor[HTML]{61CBF3}94.44 & \cellcolor[HTML]{61CBF3}86.67 & \cellcolor[HTML]{94DCF8}87.65 & \cellcolor[HTML]{CAEDFB}78.12 & \cellcolor[HTML]{61CBF3}84.54 & \cellcolor[HTML]{61CBF3}63.19 \\
Specified function name               & 33.33                         & 17.22                         & 13.99                         & 17.02                         & 9.79                          & 7.36                          \\
Extraneous elements                   & 25.93                         & 67.78                         & 56.38                         & \cellcolor[HTML]{61CBF3}93.92 & 36.60                         & 55.21                         \\ \hline
\textit{n} =                  & 162                           & 180                           & 243                           & 329                           & 194                           & 163                           \\ \hline
\end{tabular}
\end{table*}

Out of the 920 students who attempted at least one problem in B1, 916 responded to the Likert item (Ref3). Students tended towards agreeing that it was easier to solve problems by writing natural language than by writing code, with the mean response of 3.80 (\textit{SD} = 0.93). This was significantly greater than if students perceived both activities equally (\textit{t}(915) = 29.91, \textit{p} < .001, 95\% CI = 0.74, 0.86). Combined, these results suggest positive student engagement, as prompt-programming can make coding more accessible and efficient. However, students raised concerns about the depth of learning, trust in AI, and applicability to complex problems, remaining wary about relying on AI entirely.

\subsection{RQ2: Types of Mistakes in Natural Language Prompts}\label{sec:results.rq2}

There were $675$ and $821$ students who generated at least one initially incorrect prompt that required clarification in B1 and B2, respectively. Approximately half of all initial incorrect prompts were tagged for each problem. Table~\ref{fig.table_2} presents the mean and standard deviation across each problem for students who wrote an initially incorrect prompt and were coded for mistakes. Due to the samples not being independent, we were unable to test for significant differences. However, reflection on this data suggests that the problems in B2 were harder than in B1, given the lower success rate (see Table~\ref{fig.table_2}). There further appear to be more conversations and more messages sent on the third question in B1 and all of the B2 problems. Interestingly, it seems students may be making fewer mistakes in their initial prompt for the B2 problems. 

Table~\ref{fig.table_3} presents percentages of missing prompt elements out of the tagged first prompts that resulted in incorrect code. The top three most frequent mistakes are colored for each problem. Although exact argument names were not strictly required by the tool when validating code, we coded their omission as an error because the graphical input-output specification shown to students provided input variable names.  We therefore considered these part of the complete specification. In addition, good argument names supported the readability of the generated solution, and many students themselves treated them as important for producing accurate code. The results demonstrated that students were able to explicitly and consistently provide a \textbf{Specified Function Name} in their prompt. For B1 problems, missing the \textbf{Appropriate Expected Output/Return} was quite common, and refers to failing to describe what should be returned as output from the function. This code was not applicable to the B2 problems as they did not return a value (but modified inputs via reference). Common errors were omitting a \textbf{Specified Function Return Type} or \textbf{Specified Argument Names} from a prompt across most of the problems. The results suggested that omissions often co-occurred. We further investigated this through a correlational analysis presented in Table~\ref{fig.table_4}. 

\begin{table*}[t!]
  \footnotesize
  \caption{Pearson correlation coefficients between coded mistake types, reported separately for each problem among initially incorrect prompts. The $n$ column indicates the number of prompts coded with the row mistake type *\textit{p} < .05; **\textit{p} < .01 (two-tailed).}
  \label{fig.table_4}
  \begin{tabular}{llccccccccc}
  \hline
   &  & \multicolumn{8}{c}{Mistake Type Respective to Problem} & \\ \cline{3-10}
   &  & \begin{tabular}[c]{@{}c@{}}Appropriate\\ Expected\\ Output/Return\end{tabular} & \begin{tabular}[c]{@{}c@{}}Appropriate\\ Functionality\\ Explanation\end{tabular} & \begin{tabular}[c]{@{}c@{}}Appropriate\\ Required\\ Input(s)\end{tabular} & \begin{tabular}[c]{@{}c@{}}Correct\\ Order of \\ Arguments\end{tabular} & \begin{tabular}[c]{@{}c@{}}Specified\\ Argument\\ Names\end{tabular} & \begin{tabular}[c]{@{}c@{}}Specified\\ Function\\ Return Type\end{tabular} & \begin{tabular}[c]{@{}c@{}}Specified\\ Function\\ Name\end{tabular} & \begin{tabular}[c]{@{}c@{}}Extraneous\\ Elements\end{tabular} & $n$ =  \\ \cline{3-11}
   & B1-1 &  & \cellcolor[HTML]{A8DDA8}.56* & .06 & \cellcolor[HTML]{EDF7ED}.16* & -.09 & -.01 & -.10 & \cellcolor[HTML]{CFECCF}.30** & 104    \\
   & B1-2 &  & \cellcolor[HTML]{7BC67B}.74** & .07 & .12 & .10 & -.04 & \cellcolor[HTML]{CFECCF}-.31** & \cellcolor[HTML]{A8DDA8}.55** & 150    \\
   & B1-3 &  & \cellcolor[HTML]{7BC67B}.69** & .03 & .03 & .01 & -.05 & .02 & \cellcolor[HTML]{CFECCF}.23** & 217    \\
   & B2-1 &  & na & na & na & na & na & na & na & NA     \\
   & B2-2 &  & na & na & na & na & na & na & na & NA     \\
  \multirow{-6}{*}{\begin{tabular}[c]{@{}l@{}}Appropriate \\ Expected \\ Output/Return\end{tabular}} & B2-3 &  & na & na & na & na & na & na & na & NA     \\ \cline{2-11}
   & B1-1 &  &  & -.03 & .15 & -.14 & -.11 & -.10 & \cellcolor[HTML]{A8DDA8}.53** & 71     \\
   & B1-2 &  &  & .07 & .08 & .04 & -.09 & \cellcolor[HTML]{CFECCF}-.39** & \cellcolor[HTML]{7BC67B}.74** & 132    \\
   & B1-3 &  &  & -.02 & -.00 & .06 & \cellcolor[HTML]{EDF7ED}-.13* & -.06 & \cellcolor[HTML]{CFECCF}.34** & 194    \\
   & B2-1 &  &  & .07 & -.03 & .05 & -.05 & \cellcolor[HTML]{CFECCF}-.20** & \cellcolor[HTML]{4FAE4F}.85** & 307    \\
   & B2-2 &  &  & -.09 & -.20 & \cellcolor[HTML]{CFECCF}-.29** & \cellcolor[HTML]{EDF7ED}-.17* & .07 & \cellcolor[HTML]{4FAE4F}.91** & 65     \\
  \multirow{-6}{*}{\begin{tabular}[c]{@{}l@{}}Appropriate \\ Functionality \\ Explanation\end{tabular}} & B2-3 &  &  & -.18 & -.09 & \cellcolor[HTML]{CFECCF}-.26** & -.15 & -.16 & \cellcolor[HTML]{4FAE4F}.93** & 94     \\ \cline{2-11}
   & B1-1 &  &  &  & \cellcolor[HTML]{A8DDA8}.52** & \cellcolor[HTML]{A8DDA8}.52** & \cellcolor[HTML]{CFECCF}.21** & -.02 & .06 & 128    \\
   & B1-2 &  &  &  & \cellcolor[HTML]{A8DDA8}.49** & \cellcolor[HTML]{A8DDA8}.44** & \cellcolor[HTML]{A8DDA8}.51** & .10 & .05 & 133    \\
   & B1-3 &  &  &  & \cellcolor[HTML]{A8DDA8}.45** & \cellcolor[HTML]{A8DDA8}.40** & \cellcolor[HTML]{CFECCF}.30** & .06 & -.06 & 169    \\
   & B2-1 &  &  &  & \cellcolor[HTML]{CFECCF}.36** & \cellcolor[HTML]{CFECCF}.32** & \cellcolor[HTML]{CFECCF}.33** & \cellcolor[HTML]{CFECCF}.22** & .00 & 134    \\
   & B2-2 &  &  &  & \cellcolor[HTML]{CFECCF}.25* & \cellcolor[HTML]{CFECCF}.23** & \cellcolor[HTML]{EDF7ED}.16* & .19 & -.04 & 101    \\
  \multirow{-6}{*}{\begin{tabular}[c]{@{}l@{}}Appropriate \\ Required \\ Input(s)\end{tabular}} & B2-3 &  &  &  & \cellcolor[HTML]{A8DDA8}.49** & \cellcolor[HTML]{7BC67B}.62** & \cellcolor[HTML]{A8DDA8}.50** & \cellcolor[HTML]{A8DDA8}.45** & -.11 & 47     \\ \cline{2-11}
   & B1-1 &  &  &  &  & \cellcolor[HTML]{CFECCF}.31** & .08 & .04 & \cellcolor[HTML]{CFECCF}.22** & 82     \\
   & B1-2 &  &  &  &  & \cellcolor[HTML]{CFECCF}.39** & \cellcolor[HTML]{CFECCF}.25** & .08 & \cellcolor[HTML]{EDF7ED}.15* & 72     \\
   & B1-3 &  &  &  &  & \cellcolor[HTML]{CFECCF}.32** & \cellcolor[HTML]{CFECCF}.25** & -.00 & -.07 & 94     \\
   & B2-1 &  &  &  &  & .13 & \cellcolor[HTML]{CFECCF}.31** & \cellcolor[HTML]{CFECCF}.34** & .06 & 83     \\
   & B2-2 &  &  &  &  & \cellcolor[HTML]{A8DDA8}.59** & \cellcolor[HTML]{A8DDA8}.57** & \cellcolor[HTML]{A8DDA8}.45** & \cellcolor[HTML]{CFECCF}-.31** & 58     \\
  \multirow{-6}{*}{\begin{tabular}[c]{@{}l@{}}Correct \\ Order of \\ Arguments\end{tabular}} & B2-3 &  &  &  &  & \cellcolor[HTML]{A8DDA8}.59** & \cellcolor[HTML]{A8DDA8}.42** & \cellcolor[HTML]{CFECCF}.35** & -.02 & 95     \\ \cline{2-11}
   & B1-1 &  &  &  &  &  & .09 & \cellcolor[HTML]{EDF7ED}.16* & -.07 & 145    \\
   & B1-2 &  &  &  &  &  & \cellcolor[HTML]{A8DDA8}.45** & -.01 & .07 & 147    \\
   & B1-3 &  &  &  &  &  & \cellcolor[HTML]{CFECCF}.32** & -.03 & -.02 & 200    \\
   & B2-1 &  &  &  &  &  & \cellcolor[HTML]{CFECCF}.24** & \cellcolor[HTML]{CFECCF}.25** & .03 & 117    \\
   & B2-2 &  &  &  &  &  & \cellcolor[HTML]{A8DDA8}.55** & \cellcolor[HTML]{CFECCF}.34** & \cellcolor[HTML]{CFECCF}-.35** & 156    \\
  \multirow{-6}{*}{\begin{tabular}[c]{@{}l@{}}Specified \\ Argument \\ Names\end{tabular}} & B2-3 &  &  &  &  &  & \cellcolor[HTML]{A8DDA8}.48** & \cellcolor[HTML]{A8DDA8}.44** & \cellcolor[HTML]{CFECCF}-.23* & 93     \\ \cline{2-11}
   & B1-1 &  &  &  &  &  &  & .00 & -.10 & 153    \\
   & B1-2 &  &  &  &  &  &  & .09 & -.10 & 156    \\
   & B1-3 &  &  &  &  &  &  & .08 & \cellcolor[HTML]{EDF7ED}-.18** & 213    \\
   & B2-1 &  &  &  &  &  &  & \cellcolor[HTML]{EDF7ED}.16** & -.07 & 257    \\
   & B2-2 &  &  &  &  &  &  & .17 & -.14 & 164    \\
  \multirow{-6}{*}{\begin{tabular}[c]{@{}l@{}}Specified \\ Function \\ Return Type\end{tabular}} & B2-3 &  &  &  &  &  &  & \cellcolor[HTML]{CFECCF}.33* & -.12 & 103    \\ \cline{2-11}
   & B1-1 &  &  &  &  &  &  &  & .03 & 54     \\
   & B1-2 &  &  &  &  &  &  &  & \cellcolor[HTML]{CFECCF}-.38** & 31     \\
   & B1-3 &  &  &  &  &  &  &  & .04 & 34     \\
   & B2-1 &  &  &  &  &  &  &  & \cellcolor[HTML]{EDF7ED}-.16** & 56     \\
   & B2-2 &  &  &  &  &  &  &  & -.01 & 19     \\
  \multirow{-6}{*}{\begin{tabular}[c]{@{}l@{}}Specified \\ Function \\ Name\end{tabular}} & B2-3 &  &  &  &  &  &  &  & -.12 & 12     \\ \cline{2-11}
   & B1-1 &  &  &  &  &  &  &  &  & 42     \\
   & B1-2 &  &  &  &  &  &  &  &  & 122    \\
   & B1-3 &  &  &  &  &  &  &  &  & 137    \\
   & B2-1 &  &  &  &  &  &  &  &  & 309    \\
   & B2-2 &  &  &  &  &  &  &  &  & 71     \\
  \multirow{-6}{*}{\begin{tabular}[c]{@{}l@{}}Extraneous \\ Elements\end{tabular}} & B2-3 &  &  &  &  &  &  &  &  & 90     \\ \hline
  \end{tabular}
\end{table*}

There were, notably, significant correlations between the prompts failing to describe \textbf{Appropriate Required Input(s)}, \textbf{Specified Argument Names}, \textbf{Specified Function Return Type}, and a \textbf{Correct Order of Arguments}, across the majority of problems. For example, one student's prompt aiming to solve ``count negatives'' stated,
``Write a program called foo that goes through an array and outputs amount of negative numbers in the array. The inputs for foo should be an integer array variable and an integer size variable,''
which fails to describe both argument names and the function return type. The correlational evidence indicated the absence of an \textbf{Appropriate Functionality Explanation} was strongly related to the absence of an \textbf{Appropriate Expected Output/Return} and to the prompt \textit{containing} \textbf{Extraneous Elements}. Moreover, co-occurrences appeared to often resolve simultaneously. Pearson correlation coefficients were calculated by problem between the number of messages and number of mistakes. The results were non-significant, B1-2 (\textit{r} = .01), B1-3 (\textit{r} = -.02), and B2-3 (\textit{r} = .10), or weak, B1-1 (\textit{r} = .19, \textit{p} < .05), B2-1 (\textit{r} = .17, \textit{p} < .01), and B2-2 (\textit{r} = .17, \textit{p} < .05), indicating little evidence to support the number of mistakes influencing the number of prompts needed.

\subsection{RQ3: Student Strategies for Refining Prompts}\label{sec:results.rq3}

\begin{figure*}[t!]
\centering
	\includegraphics[width=\linewidth]{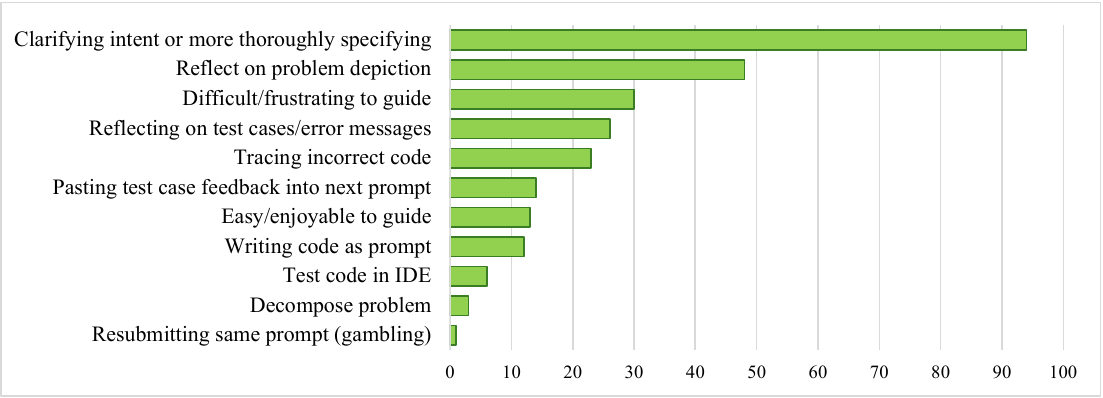}
    \vspace{-7mm}
     \caption{Strategies reported to refine prompts following unsuccessful attempts on the second batch of problems (B2) (\textit{n} = 174).}
     \Description[RQ3 coded strategies for refining prompts after incorrect AI output]{Horizontal bar chart showing coded strategies from student reflections on the second batch of Prompt Problems, based on 174 coded responses. The x-axis shows the number of coded responses, ranging from 0 to 100, and the y-axis lists eleven strategies. The strategies, ordered from least to most frequent in the chart, are Resubmitting same prompt, labeled as gambling; Decompose problem; Test code in IDE; Writing code as prompt; Easy/enjoyable to guide; Pasting test case feedback into next prompt; Tracing incorrect code; Reflecting on test cases or error messages; Difficult/frustrating to guide; Reflect on problem depiction; and Clarifying intent or more thoroughly specifying. The most frequent strategy is clarifying intent or specifying the prompt more thoroughly, followed by reflecting on the visual problem depiction. Moderately frequent strategies include reporting difficulty or frustration when guiding the AI, reflecting on test cases or error messages, tracing incorrect code, and pasting test case feedback into the next prompt. Less frequent strategies include writing code directly as the prompt, testing code in another IDE, decomposing the problem, and resubmitting the same prompt.}
    \label{fig.barchart}
\end{figure*}

When their first prompt submission in B2 was not successful, students reported using a variety of strategies to guide the AI model to a correct solution (Figure~\ref{fig.barchart}). Out of the students who attempted B2, made at least one mistake, and responded to Ref4, 174 were coded for strategies. The most frequently identified strategy involved \textbf{Clarifying intent or more thoroughly specifying} details in their prompt. Many students also indicated they would further \textbf{Reflect on problem depiction} itself. These two themes commonly went together, for example, one student reflected, ``I tried to give more and more conditions in detail. I paid most attention to the input and the limitations of the input in regards to output.'' 

There are three key sources of information for students to inspect when the generated code was incorrect: the graphical problem depiction, the generated code, and the failing test cases. Students reported using all of these sources to refine their prompts when the generated code was faulty. The graphical problem depiction was the most commonly used (\textbf{Reflect on problem depiction}). For example, one student stated ``Reading the scenarios again to understand what we are asking for helps me to reconstruct an appropriate demand.'' \textbf{Reflecting on test cases/error messages} and \textbf{Tracing incorrect code} were less likely to be reported as strategies. A student commented, ``For identifying the problem, I would look at code for errors based off what the error is.'' A few students mentioned attempting to break the problem down into smaller parts (\textbf{Decompose problem}). Several reported that they tested the generated code in other environments before submitting (\textbf{Test code in IDE}). 

A smaller number of students wrote about approaches less conducive to the intention of the learning environment, such as \textbf{Pasting test case feedback into next prompt} directly or \textbf{Writing code as prompt}. For example, \begin{quote}Some of the strategies I used were simply just feeding in the inputs and desired outputs from the test cases where the AI model was mostly able to get it correct. If the AI model was not correct, I would indicate which test case it failed, copy and pasting the inputs and the expected outputs vs what the code generated, and the AI model was mostly able to fix itself with ease.\end{quote}

Although one might assume repeatedly resubmitting the same prompt to see if the model would generate correct code to be a negative strategy, i.e., \textbf{Resubmitting same prompt (gambling)}, the one case we found was the opposite. The student outlined learning from observing how different the code could be from prompt to prompt. They wrote, \begin{quote}I initially tried resubmitting the same response, because the code would be different with every submission. \ldots I paid most attention to the different outputs it created by me submitting the same paragraph of information, as it gave me an idea of what the AI wasn't understanding in my description.\end{quote}

Some responses were off-topic but still insightful, such as the AI being \textbf{Difficult/frustrating to guide}, specifically, ``This one was more tedious and pedantic as it got more complicated. By having to state inputs or missing out on any information would screw up the output code \ldots felt like it needed more work than just writing the code itself." Another commented, ``It does get really frustrating when I have to paraphrase my demand many, many times and be very specific''. Many of the frustrations referred to feeling misunderstood by the model. Fewer students identified the process of solving B2 tasks as \textbf{Easy/enjoyable to guide}. 

Notably, for each strategy, there was no significant relationship with problems solved in B2, except for \textbf{Writing code as prompt} (\textit{r} = .20, \textit{p} < .05), which should be treated cautiously given that only twelve students were coded using this. 



\section{Discussion}
\label{sec:discussion}

\subsection{Perceived Ease and Vibe Coding}

The work aimed to help students develop their capabilities in using AI for generating accurate code \cite{ebert2023generative}. Further, the tool design was theoretically grounded with the intention of reducing students' cognitive load by removing the effort of writing correct syntax, freeing up working memory to focus on higher-level problem-solving and comprehension skills \cite{richmond2025benefits, risko2016cognitive}. We provided evidence that, for many students, the tasks are working as intended. Students drew attention to the tasks supporting problem-solving and noted the relationship with code comprehension. The ability to abstract from a few concrete examples to the general purpose, a requirement of these Prompt Problems, is a  fundamental challenge in CS known as ``Generic Programming,'' proposed by Alan Kay in 1981 ~\cite{kay1981generic, musser1988generic}. Some students noted understanding the problem as the most difficult aspect of programming, reinforcing our study's significance. Further, students clearly identified \textbf{Learning AI/Prompting Skills} as a notable part of these exercises. With no formal instruction on how to interact with AI for code production, students felt they were developing a new, industry-critical skill ~\cite{ebert2023generative}. Together, we have provided a baseline for what students feel they learn when engaging with dialogue-based prompting compared to traditional coding, and set the stage for future research to determine effective instruction around this process.  

We have provided preliminary evidence that these exercises can free up cognitive resources and make the problem-solving process more efficient, while also being enjoyable. Among the most frequently reported perceptions was that these were \textbf{Easier/Simpler/Faster}, supported by the quantitative findings, as well as \textbf{Fun/Interesting}. Evidence from educational psychology overwhelmingly supports the influence of positive experiences on achievement, retention, motivation, engagement, and more \cite{mega2014makes, pekrun2006control, pekrun2014introduction, pekrun2017achievement, schukajlow2017emotions, usher2009sources}, highlighting the value of these tasks to students' pursuit of computing education. While it may seem intuitive that writing natural language prompts is preferable, other themes clearly indicated that this was not a universal truth (e.g., \textbf{Unhelpful for Learning/Thinking/Programming}), and these perspectives should be further investigated.

We note \emph{perceived} ease may not reflect the \emph{actual} effort needed to successfully produce working code. Vaithilingam et al. found that for experienced programmers, producing blocks of code using GitHub Copilot is faster than searching for solutions online, but the generated code is often incorrect or requires substantial modification \cite{vaithilingam2022expectation}. However, participants preferred using Copilot because it provided a useful starting point, despite often encountering challenges in understanding and debugging the AI-generated code, which sometimes led to longer debugging sessions than in traditional coding approaches. In an educational context, and especially as programming tasks grow in complexity, students may initially feel that natural language prompts simplify programming, but this may not align with the actual effort required to iteratively refine faulty AI-generated solutions. Interestingly, our study found a conflict in the data on the perceived reliability of AI-generated code. Some students believed AI code to be accurate and efficient, while others noted it can produce incorrect solutions, does not work for complex problems, and they preferred traditional coding. A concern for the latter group of students is that they will not gain the advantages from the tasks, as a lack of trust in the tool can prevent cognitive offloading \cite{peng2025cognitive}. One design implication that should be empirically investigated is including the student suggestion of a combined approach of prompting and manually editing the code. 

\looseness-1Of course, even faulty AI-generated code can be fed back to an AI model for refinement. This idea underpins the emerging phenomenon of ``vibe coding''\footnote{\url{https://en.wikipedia.org/wiki/Vibe_coding}}, where users describe a problem in natural language and allow an AI model to generate the implementation. Vibe coding shifts the focus toward guiding and refining AI outputs without necessarily comprehending the underlying code. Errors can be fed back to the AI model repeatedly and any resulting code changes accepted without critique. Students in our study commented on using this strategy for prompt-refinement. While vibe coding may serve as a hobbyist approach to interacting with code, it does not meet the educational goals of a scientific discipline and is an example of problematic cognitive offloading ~\cite{gerlich2025ai}. We want students to see programming as both a way of expressing computational thinking and problem-solving skills, and to develop foundational knowledge of programming principles required for code understanding and debugging. The activity presented in this paper emphasizes natural language coding, but also reveals the generated code and the failed test cases, which can be used when refining prompts. Tracing incorrect code was reported by some students as a strategy for recovering from errors, as was reflecting on the test cases and error messages. Future work should consider greater scaffolding to support appropriate prompt-refinement approaches.   

\subsection{Scaffolding the Inclusion of Key Information}
\looseness-1Analyzing students' initially unsuccessful prompts, we found that failure to describe the intention of the code was related to omitting the correct expected output/return and to including irrelevant information. This logically follows from a misunderstanding of the problem itself. We frequently observed omissions of critical details~\cite{prather2024widening}, such as function return types and argument names. It is highly probable that students tended to rely excessively on the AI's ability to infer missing elements or simply did not acknowledge their importance (common among non-experts~\cite{macneil2021framing}), resulting in ambiguous prompts and incorrect or incomplete code. This kind of error-making seems to align with findings on developers ~\cite{nam2025prompting}, indicating where educators may want to provide explicit instruction. Again, excessive reliance on AI can result in \textit{too much} cognitive offloading and negatively impact critical thinking skills~\cite{gerlich2025ai}. Novice programmers still need to retain a level of engagement that does not result in critical omissions from their prompts or the resulting code. 

With a view to controlling the level of cognitive offloading, possible design implications to explore include three structured methods to systematically encourage students to include the relevant information in their prompts. First, adapting the Parsons problem technique could explicitly train students to differentiate between relevant and irrelevant information \cite{denny2008evaluating}. Specifically, in Parsons problems, students order blocks of code and the blocks available for selection often include `distractors,' which are irrelevant to the correct solution. Parsons problems effectively reduce cognitive load while reinforcing problem-solving skills~\cite{ericson2022parsons}. 
Drawing on this, students may construct prompts by assembling predefined fragments (essential or extraneous) into a coherent, correctly ordered prompt. More challenging scenarios could be progressively introduced by adding distractor fragments, encouraging students to critically evaluate each component's necessity. Given the empirical evidence for common missing elements, another approach could be structured prompt templates, which
guide students in specifying necessary elements, such as argument names, required inputs, and return types. Such templates might initially provide significant structure, then gradually reduce support. 
Finally, an AI middleware component could analyze prompts for completeness before generating code, alerting students immediately to missing details or offering hints for scaffolding. This mechanism may help students internalize the importance of prompt completeness, preparing them to handle more complex tasks independently. These suggested strategies aim to progressively enhance students' prompt engineering skills and encourage more thoughtful interactions with GenAI systems.

\looseness-1An important outcome from this paper is baseline information on the ways novices interact with dialogue-based prompting without prior instruction. We should expect students to engage with iterative refinement, even on simple problems. Notably, we provided preliminary evidence that more initial errors do not require greater effort for students to resolve. Our work suggests that common and related mistakes pertained to excluding details from the prompts, such as explicitly identifying inputs, argument names, and types. Understanding that they were providing an insufficient level of detail could potentially explain why students appeared to resolve these issues simultaneously. This may be another potential benefit of cognitive offloading, specifically, students taking a high-level view of the patterns of their mistakes, rather than linearly debugging. The findings cue future work on the interrelatedness of error-making and how natural language prompts can cause and resolve multiple mistakes together. There could be further value in students reflecting on how multiple errors stem from their specific prompting approach. 

\subsection{Refining Prompts: Strategies and Challenges}

\looseness-1The most commonly reported strategy for refining prompts in this study was further clarifying or specifying. Using a tool that supported dialogue-based interactions allowed students to refine their requests iteratively rather than requiring a single prompt that included all necessary details, contrasting with earlier work on prompt-based programming ~\cite{denny2024prompt}. Both approaches have merit: dialogue-based interactions more closely mirror real-world AI-assisted development practices, whereas the one-shot approach requires deeper initial thinking about problem specifications. However, it is possible many students took a brute-force approach to prompt refining using dialogue-based interactions (i.e. adding prompt details only as they surfaced through feedback). The extensive cognitive offloading using this approach to iterative refinement may have negative consequences ~\cite{gerlich2025ai}. Future research should explore how to best integrate or sequence these two methods, with a view to optimizing information in students' working memories.

While reconsidering the problem after a failed prompt is useful, we did not see as much evidence that students were attending to other sources that may inform where the failure could be originating (i.e., generated code; test cases). The ability to carefully read and analyze code is a core skill, one that should not be replaced by prompt programming. 
Thus, the design of AI-assisted programming tools should include scaffolding to promote code comprehension. This could be done by 
requiring students to reflect on discrepancies between the generated code and the failing test cases before revising their prompt.
A more radical approach could include deliberately injecting bugs that require manual detection and fixing, ensuring the generated code is carefully studied. 

\looseness-1Many students expressed significant frustration when refining their prompts, specifically centered around difficulty in communicating with the AI and feeling misunderstood by the model, which could be detrimental ~\cite{peng2025cognitive}. This frustration likely stems from the limitations of current LLMs in correctly interpreting human intent, and may be alleviated as model capabilities continue to improve. Nevertheless, frustration in problem-solving can be beneficial to the learning process ~\cite{riegel2021frustration, margulieux2025biological} and the iterative process of interacting with AI is a core part of learning effective prompt engineering ~\cite{ebert2023generative}. Further work is needed in encouraging students to engage with the prompt-refining strategies most beneficial to their overall computing education, while not introducing significant technological frustrations.

\subsection{Limitations}
Our study has several limitations. The findings may not directly generalize to contexts outside of introductory-level C programming tasks. 
The problems we used were relatively simple 
and may not fully capture the complexities and challenges students face when using these tools in more realistic or advanced programming scenarios. Future work might conduct an in-depth analysis of the iterative refinement of student prompts and see how this aligns with the strategies they report using and overall effectiveness. This may be particularly interesting in more advanced courses where test cases are not provided.
Future research would benefit from explicit evaluation of cognitive load, to provide insight beyond theoretical alignment~\cite{ouwehand2021measuring}. 
Under the terms of our ethics approval, we were unable to access student demographic data, necessitating future research on whether this approach is equitable. 
Without providing specific guidelines on how students should interact with the AI, some results (such as messages sent) should be treated cautiously, as different students may have taken a more casual approach to their attempts at the problems. 
Due to the exploratory study design, we lacked independent samples to conduct comparative analyses. 
Further, our study lacks a direct A/B evaluation between dialogue-based and single-shot prompting. Without this comparison, we cannot determine whether the dialogue-based approach used in this study meaningfully improves student experiences or outcomes compared to alternative prompting approaches. 



\section{Conclusion}
\label{sec:conclusion}
We explored introductory programming students' experiences, mistakes, and strategies when engaging with natural language, dialogue-based prompt programming tasks. Through analyzing students' mistakes, error-correction strategies, and reflections, we found that they generally perceived natural language prompting as beneficial and easier than traditional coding, highlighting reduced syntax-related cognitive load and enhanced engagement with computational problem-solving. However, students also struggled with clearly specifying function requirements in their initial prompts. When encountering errors, students typically refined their prompts through iterative clarifications, frequently returning to the problem depiction. They less often attended to strategies beneficial to their learning, such as tracing generated code and reflecting on test cases and errors. A promising area for future work is exploring effective pedagogical methods and scaffolding techniques to help students develop stronger natural language prompting skills. Investigating tools and instructional strategies that provide interactive, immediate feedback on prompt quality and completeness could significantly improve students' learning outcomes. Furthermore, examining how students' prompting skills evolve over time in longitudinal studies and how these skills transfer to other programming contexts would provide deeper insights into integrating Prompt Problems effectively within broader computing curricula. 


\begin{acks}
This work was supported by Research Council of Finland grant \#356114. Funded/Cofunded by the European Union (ERC, TOPS, 101039090). Views and opinions expressed are however those of the author(s) only and do not necessarily reflect those of the European Union or the European Research Council. Neither the European Union nor the granting authority can be held responsible for them.
\end{acks}

\bibliographystyle{ACM-Reference-Format}
\balance
\bibliography{main}

\end{document}